\documentclass[%
preprint,amsmath,amssymb,aps,pra,floatfix]{revtex4-2}

\usepackage{graphicx}
\usepackage{dcolumn}
\usepackage{bm}

\usepackage{xcolor} 
\usepackage{float}
\usepackage{amsmath, amsthm, amssymb}
\newtheorem{corollary}{Corollary}[section]
\newtheorem{lemma}{Lemma}

\begin{document}
%\linenumbers

\title{Segal-Bargmann type spaces related to non-rotational measure, and entanglement of bipartite squeezed coherent states
}

\author{K. G{\'o}rska}
\affiliation{H. Niewodnicza{\'n}ski Institute of Nuclear Physics, Polish Academy of Sciences, ul.Eljasza-Radzikowskiego 152, 31-342 Krak{\'o}w, Poland.}
%\orcid{0000-0003-4798-9146}
%\email{katarzyna.gorska@ifj.edu.pl}
\author{A. Horzela}
\affiliation{H. Niewodnicza{\'n}ski Institute of Nuclear Physics, Polish Academy of Sciences, ul.Eljasza-Radzikowskiego 152, PL 31342 Krak{\'o}w, Poland.}
%\orcid{0000-0001-8098-2275}
%\email{andrzej.horzela@ifj.edu.pl}
\author{D. Kołaczek}
\affiliation{Department of Applied Mathematics, University of Agriculture in Kraków, ul. Balicka 253c, 30–198 Kraków, Poland}
%\orcid{0000-0001-9840-8829}
%\email{Damian.Kolaczek@urk.edu.pl}
\author{B. J. Spisak}
\email[]{bjs@agh.edu.pl}
\affiliation{AGH University of Krakow, Faculty of Physics and Applied Computer Science, \\ al. A. Mickiewicza 30, 30-059 Krakow, Poland}
%\orcid{0000-0001-6716-5182}
\author{F. H. Szafraniec}
\affiliation{Instytut Matematyki, Uniwersytet Jagiello{\'n}ski, ul. {\L}ojasiewicza 6, 30 348 Krak{\'o}w, Poland.}
%\orcid{0000-0003-1957-1658}
%\email{umszafra@cyf-kr.edu.pl}

\begin{abstract}
Entanglement of bipartite squeezed states generated by holomorphic Hermite functions of two complex variables is investigated using phase-space approach based on the Wigner distribution function. 
Orthogonality of the holomorphic Hermite functions implies the relationship between certain real parameter associated with the non-rotational measure in the Bargmann space and the squeezing parameter.
The mutual relation between squeezing and entanglement is elucidated with the help of Peres-Horodecki positive partial transpose criterion formulated in the phase-space version for continuous-variable systems. 
The quantitative characteristics of the entanglement is determined using the log-negativity criterion.
The oscillator-like model of a two-particle quantum-mechanical system is developed to illustrate the presented findings.
\end{abstract}
\maketitle

%%%%%%%%%%%%%%%%%%%%%%%%%%%%%%%%%%%%%%%%%%%%%%%%%%%%%%%%%%%%%%%%%%%%%%%%%%%%%%%%%%%%%%%%
\section{\label{sec:1}Introduction}
%%%%%%%%%%%%%%%%%%%%%%%%%%%%%%%%%%%%%%%%%%%%%%%%%%%%%%%%%%%%%%%%%%%%%%%%%%%%%%%%%%%%%%%%
Coherent states have been the subject of intensive research for decades, and they are one of the central issues in modern quantum theory~\cite{Zhang_RMP62p867y1990,JPGazeau09,JRKlauder85}.
One of the methods of their introduction is minimizing the Heisenberg uncertainty relation~\cite{Schleich_book2001,Gerry_book2004}.
As a result, the wave function of the coherent state is given by the formula
\begin{equation}
\psi_{\rm{coh}}(x; x_0, p_0)
=
\sqrt[4]{\frac{1}{2\pi\Delta x}}
\exp\left[-\left(\frac{x-x_0}{2\Delta x}\right)^2
+
\frac{i}{\hbar}p_0 \Delta x
\left(\frac{x}{\Delta x}-\frac{x_0}{2\Delta x}
\right)
\right],
\label{eq:20230123-1}
\end{equation}
where $x_0$ corresponds to the point of localization of this wave function in the real space, and $p_0$ is its initial momentum, and $\Delta x$ is the width of the Gaussian, such that $\Delta x\Delta p=\hbar/2$. 

Alternatively, the coherent state can be generated by the Glauber operator, $\hat{D}(1;z_0)$, acting on the vacuum state, $\left|\phi_0\right\rangle$,  according to the equation $\left|z_0\right\rangle=\hat{D}(1;z_0)\left|\phi_0\right\rangle$, where $\hat{D}(1;z_0)=\exp\left[\overline{z}_0\hat{a}-z_0\hat{a}^{\dagger}\right]$, $z_0\in\mathbb{C}$,  and the bar-symbol denotes complex conjugation,  $\hat{a}$, $\hat{a}^\dagger$ are harmonic oscillator annihilation and creation operators, respectively,  wherein the vacuum state is defined by the equation $\hat{a}\left|\phi_0\right\rangle=0$. 
Let us note that the coordinate (Schr{\"o}dinger) representation of the coherent state $\left|z_0\right\rangle$ generated by the Glauber operator,
\begin{equation}
\left\langle x|z_0\right\rangle
=
\left\langle x\right|\hat{D}(1;z_0)\left|\phi_0\right\rangle
=
\psi_{\rm{coh}}(x;z_0),
\label{eq:20230123-2}
\end{equation}
leads to the wave function given by Eq.~(\ref{eq:20230123-1}), however with the complex parameter $z_0$ in the form $z_0 = x_0/\Delta x + i p_0\Delta x/(\sqrt{2}\hbar)$ which is interpreted as the localization point of the coherent state in the phase space plane. 
Expansion of the wave function of the coherent state $\psi_{\rm{coh}}(x;z_0)$ in Hermite's function basis~\cite{Arfken},  
\begin{equation}
\phi_n(x) 
= \frac{1}{\sqrt{n!}}\sqrt[4]{\frac{1}{2^{2n+1}\pi \Delta x}} 
H_n\left(\frac{x}{\Delta x}\right) \exp\left[-\frac{x^2}{2(\Delta x)^2}\right],
\label{eq:20230123-4}
\end{equation}
leads to the following result
\begin{equation}
\psi_{\rm{coh}}(x;z_0)
=
e^{-|z_0|^2/2}
\sum_{n=0}^{\infty}
\frac{z_0^n}{\sqrt{n!}}
\phi_n(x).
\label{eq:20230123-3}
\end{equation}
The coherent state wave function given by Eq.~(\ref{eq:20230123-3}) has a convenient form for further generalization. 
Namely, the action of the single-mode squeeze operator, $\hat{S}(1;\xi)=\exp\left[(\xi\hat{a}^{\dagger 2}-\overline{\xi}\hat{a}^2)/2\right]$, on the coherent state represented by the wave function $\psi_{\rm{coh}}(x;z_0)$ generates the wave function of the squeezed state, $\widetilde{\psi}_{\rm{coh}}(x; z_0, \xi)$ which is called the squeezed coherent state wave function, where the number $\xi\in\mathbb{C}$ is called the squeezing parameter~\cite{MOSculy1997}. 
Referring to Eq.~(\ref{eq:20230123-3}), this wave function can be expressed in a different orthogonal basis, $\{\chi_q(x;\xi)\}_{q=0}^{\infty}$, which is a consequence of the action of the squeeze operator on elements of the basis $\{\phi_q(x)\}_{q=0}^{\infty}$ according to the unitary transformation $\left\langle x |\chi_q(\xi)\right\rangle = \left\langle x \right|\hat{S}(1;\xi)\left|\phi_q\right\rangle$.
In this way, we generalized Eq.~(\ref{eq:20230123-3}) to the following form
\begin{equation}
\widetilde{\psi}_{\rm{coh}}(x; z_0, \xi)
=
e^{-|z_0|^2/2}
\sum_{q=0}^{\infty}
\frac{z_0^q}{\sqrt{q!}}
\chi_q(x;\xi).
\label{eq:20230307-00}
\end{equation}
Simultaneously, let us note that the unitary transformation of the orthogonal basis elements holds the condition of completeness~\cite{JPGazeau09,Klauder1,Klauder2,JKlauder63}
\begin{equation}
\int_{\mathbb{C}}|z_0,\xi\rangle\langle z_0,\xi|\frac{dz_0}{\pi}
=
\sum\limits_{q=0}^{\infty}|\chi_q\rangle\langle \chi_q|
=
\textrm{Id}_{\mathcal{F}},
\label{eq:20230307}
\end{equation}
where the symbol ${\textrm{Id}}_{\mathcal{F}}$ denotes the identity operator over the Fock space $\mathcal{F}$~\cite{KRParthasarathy92}. 
It is worth emphasizing that the fulfillment of the completeness condition~(\ref{eq:20230307}) opens the possibility to interpret coherent states as a continuous basis used to represent the quantum states and characterised by i) continuity in the label $\xi$, ii) normalizability and iii) (over)completeness. 
Properties i)--iii) have been postulated as a cornerstone of the general construction of the coherent states, which arose from Klauder's seminal papers of the early sixties~\cite{Klauder1,Klauder2}.

The presented description summarizes the main results of the one-mode picture of the squeezed coherent state construction. 
Applications and the role played by coherent states in quantum optics are described in Ref.~\cite{JPGazeau19,Dodonov_JOBQSO4pR1y2002}. 
Other applications of the coherent states cover wide range of mathematical physics issues, starting from quantization~\cite{JPGazeau09,JPGazeau22}, through cosmology~\cite{HBergeron24,HBergeron24a}, to field theory~\cite{JPGazeau23}.

Another look at the coherence states and their properties comes from the multi-mode perspective. 
However, this exciting problem states a different thread we do not develop here because it is out of the scope of the proposed research. 
So, instead of this, we focus on the simplest case; namely, we consider the two-mode problem. 
It is admittedly more straightforward, but still offers the possibility of a deeper insight into the entanglement phenomena and the role of quantum coherence. 
Moreover, in some cases, the two-mode system seems to be an intriguing research playground for testing the concepts mentioned above. 
Recently, this issue has been examined regarding the factorization of bipartite Gaussian wave functions through the canonical transformation of position and momentum coordinates~\cite{KGorska23}.
Admittedly, the authors demonstrated that this transformation can be used to factorize the wave functions under investigation. 
Still, their conclusions clearly show that the problem of entanglement of the bipartite states corresponding to these wave functions remains unsolved, because the mentioned transformation mixes degrees of freedom of the input subsystems on which they are originally defined.  
These results are generally consistent with the findings of Thirring et al.~\cite{Thirring_EPJD64p181y2011} and de la Torre et al.~\cite{Torre_EJP31p325y2010}, that a problem of state separation or their entanglement depends on how observables are assigned to subsystems and how physical subsystems are identified. 
In other words, the discussed disentanglement of states resulting from the factorization of the families of the bipartite Gaussian functions in the previous work~\cite{KGorska23} is rather formal because the unitary transformation generated by the point transformation does not preserve the decomposition into subspaces associated with the input subsystems.

The structure of the bipartite states realized with the help of the squeezed coherent states opens a way to study the quantum states, containing entanglement and coherence on an equal footing. 
In other words, these two phenomena coexist in such a state through its construction. 

The issue how to build bipartite squeezed states has been considered for many years, i.e. \cite{Schumaker_PR135p317y1986,SMBarnett87,RFBishop88,AVourdas92,AKEkert89,FHong-yi90,GYeoman93,FHong-yi96,FBagarello22}.
The construction of the bipartite squeezed coherent states is based on applying the squeeze operator $\hat{S}(1,2;\xi)$ to the two-mode vacuum state, which is introduced as the tensor product of two single-mode vacuum states, $|\phi_0(j)\rangle$ with $j=1,2$ denoting modes, namely
\begin{equation}
\widetilde{\psi}(1,2;\xi)\rangle
=
\hat{S}(1,2;\xi)\left[|\phi_0(1)\rangle\otimes|\phi_0(2)\rangle\right],
\label{eq:20230406-0}
\end{equation}
The action of the squeeze operator on such constructed state can be twofold. 
Namely, in the first case, each of these vacuum modes is separately squeezed $\hat{S}_{1}(1, 2; \xi)=\hat{S}(1;\xi)\otimes \hat{S}(2;\xi)$, where $\hat{S}(j;\xi)=\exp{[(\xi\hat{a}_j^{\dagger 2}-\overline{\xi}\hat{a}_j^2)/2]}$ for modes $j=1,2$.
In the second case, both modes are simultaneously squeezed $\hat{S}_2(1,2; \xi)=\exp{[\xi \hat{a}_1^\dagger \hat{a}_2^\dagger-\overline{\xi}\hat{a}_1\hat{a}_2]}$. 
This is a consequence of the construction of the squeeze operator, which has two inequivalent representations as an element of the $\mathfrak{su}(1,1)$ algebra~\cite{RFBishop88}.
Mutual relation between the single-mode squeezing operator $\hat{S}(\cdot; \xi)$ and the two-mode squeezing operator $\hat{S}_{2}(1, 2; \xi)$  can be written as follows~\cite{DChruscinski2004} 
\begin{equation}
\hat{S}_2(1,2;\xi)
=
\hat{R}\left(1,2;\frac{\pi}{4}\right) \left(\hat{S}(1;\xi)\otimes \hat{S}(2;-\xi)\right) \hat{R}\left(1,2;-\frac{\pi}{4}\right),
\label{eq:20230406-01}
\end{equation}
where $\hat{R}(1,2;\vartheta)$ is the mixing operator expressed by the formula~\cite{Schumaker_PR135p317y1986}
\begin{equation}
\hat{R}(1,2;\vartheta)
=
\exp\left[\vartheta (\hat{a}_1\hat{a}_2^\dagger-\hat{a}_1^\dagger\hat{a}_2)\right],
\label{eq:20230406-02}
\end{equation} 
where $\vartheta$ is the real parameter.
Then, depending on how the two-mode vacuum state, $|\phi_0(1)\rangle\otimes|\phi_0(2)\rangle$, has been squeezed we obtain two different squeezed states, i.e. 
\begin{equation}
|\widetilde{\psi}_k(1,2;\xi)\rangle
=  
\hat{S}_k(1,2;\xi)\left[|\phi_0(1)\rangle\otimes|\phi_0(2)\rangle\right]
\label{eq:20230425-1}
\end{equation}
for the index $k=1,2$, which corresponds to each of the mentioned-above methods of the state squeezing.
The action of the displacement operator, $\hat{D}(1,2;z_1,z_2)=\hat{D}(1;z_1)\otimes\hat{D}(2;z_2)$, on the obtained squeezed states, $|\widetilde{\psi}_k(1,2;\xi)\rangle$, leads to the formation of the bipartite squeezed coherent states indexed by $k$, namely
\begin{equation}
|\widetilde{\psi}_{k,\rm{coh}}(1,2;z_1,z_2, \xi)\rangle
=
\hat{D}(1,2;z_1,z_2)|\widetilde{\psi}_k(1,2;\xi)\rangle.
\label{eq:20230406}
\end{equation}
The Schr{\"o}dinger's wave functions, $\widetilde{\psi}_{k, \rm{coh}}(x_1,x_2;z_1,z_2, \xi):=\langle x_1,x_2|\widetilde{\psi}_{k,\rm{coh}}(1,2;z_1,z_2, \xi)\rangle$, where $x_1$ and $x_2$ are positions of the particles, corresponding to these bipartite squeezed coherent states given by Eq.~(\ref{eq:20230406}) above have different properties inherited from the presented construction. 
Taking into account the different nature of the squeezing operators, we suppose that such introduced wave functions encode useful features for studying entanglement phenomena in bipartite systems. 
This issue especially seems attractive in the possible interrelation between entanglement, coherence, and squeezing occurring in these systems. 
One of the ways of exploring the coexistence of the aforementioned phenomena in bipartite systems can rely on analysing the wave functions of these states in different representations. 
Usually,  the choice of such representation is justified on pragmatical grounds. 
In the considered case, using the Bargmann representation is the most promising to describe the coherent states.  
It also enables us to use methods of the analytic functions~\cite{BCHall2000,AVourdas2006}.

In the seminal work of Bargmann \cite{VBargmann61}, it was demonstrated that multimode coherent states can be generated using a product monomial basis in the Hilbert space of entire functions with a rotationally invariant measure, which is currently designated as the  Bargmann space.
In Ref.~\cite{SLLvanEijndhoven90}, a new space of holomorphic functions of one variable with a non-rotationally invariant measure, characterized by a single real parameter within the interval $(0,1)$, was introduced.
The generalization of their results to two dimensions presented in Ref.~\cite{lumini,Gorska2018} enabled us to identify this space as the Reproducing Kernel Hilbert Space (RKHS), in which the measure is also not rotationally invariant.
In the most straightforward case of the discussed construction, the absence of rotational invariance is also parameterized by the real parameter $\alpha$. 
Owing to the existence of RKHS, one can rigorously justify the completeness relations (decomposition of unity) for the basis created from holomorphic Hermite polynomials~\cite{lumini}.
Moreover, in Ref.~\cite{lumini}, the authors demonstrated that rescaling the introduced Hermite polynomials with help functions dependent on this parameter leads to the Bargmann space in which the rotational invariance of the measure is restored, and simultaneously, the basis is created from the Hermite functions dependent on the parameter $\alpha$.
These basis functions are used in the construction of generalized coherent states.
By performing the Segal-Bargmann transformation with these functions to the position space, we obtain Gaussian wave functions that minimize the uncertainty principle and which may not have a factorized form.
The above-mentioned unitary transformation in the Bargmann space transforms one basis into another. 
It is interpreted in the Schr{\"o}dinger representation as squeezing of the Gaussian wave functions.
However, it is essential to note that performing this operation on the Gaussian functions raises the question of the connection between squeezing and entanglement.
This question appears particularly pertinent given the theorem proposed by de Gosson, which states that an appropriate symplectic transformation can separate each Gaussian state~\cite{Gosson_LMP111p73y2021}.
Motivated by this observation, we perform an analysis of the entanglement of the squeezed bipartite Gaussian functions generated from a base of holomorphic Hermite functions using the Peres-Horodecki criterion expressed in the language of phase space~\cite{Simon_PRL84p2726y2000}.

The novelty of the presented approach is that it examines the squeezing and entanglement phenomena starting from the description of bipartite states in the Bargmann space. 
The first step enables us to establish a direct connection between the parameter, $\alpha$, associated with the non-rotational measure in the Bargmann space and the squeezing parameter. 
By crossing with the calculations to the phase space and applying the results of Simon and de Gosson, we associate the phenomena of squeezing and entanglement. 
Moreover, we quantify the degree of entanglement using the log-negativity criterion using the symplectic eigenvalues~\cite{Gosson_QHA2021} of the covariance matrix modified by Simon~\cite{Adesso2004}. 
Our result explicitly depends on the parameter $\alpha$, thereby affirming the connection between the parameter associated with the non-rotational measure in the Bargmann space and entanglement. 
Finally, we extend the model presented in the previous work~\cite{KGorska23} to the case of two coupled quantum-mechanical anisotropic harmonic oscillators, with the coupling and anisotropy originating directly from the squeezing parameter. 
This result complements the study of applying coupled harmonic oscillators to investigate the problem between entanglement and squeezing in continuous-variable Gaussian systems~\cite{Li_SR13p11722y2023}.
Concurrently, it refers to the examination of the two coupled harmonic oscillators in the Segal-Bargmann space representation presented in Ref.~\cite{Alonso_EPJD77p43y2023}.

The rest of this paper is organized as follows. 
In Sec.~\ref{sec:2}, we recall selected properties of the holomorphic Hermite functions and discuss their utility in constructing single- and bipartite coherent and squeezed states. 
The main results of our study and their discussion are presented in Sec.~\ref{sec:3}.
In the first part of this section, we use the phase-space approach based on the Wigner distribution function to quantitatively investigate the entanglement issue in continuous-variable systems. 
In turn, in the second part, we develop a theoretical model of a two-particle mechanical system illustrating details of our study. 
Section~\ref{sec:5} contains a summary and conclusions.
Finally, the paper concludes with an appendix containing the proofs of the Mehler formulae.
%%%%%%%%%%%%%%%%%%%%%%%%%%%%%%%%%%%%%%%%%%%%%%%%%%%%%%%%%%%%%%%%%%%%%%%%%%%%%%%%%%%%%%%%
\section{\label{sec:2}Mathematical preliminaries}
%%%%%%%%%%%%%%%%%%%%%%%%%%%%%%%%%%%%%%%%%%%%%%%%%%%%%%%%%%%%%%%%%%%%%%%%%%%%%%%%%%%%%%%%
Let ${\mathcal{H}_{\rm hol,2}}(\mathbb{C}^2,\mu(d w_1, d w_2))$ is the Bargmann space of holomorphic functions with the measure $\mu(d w_1,d w_2)$ in the standard Gaussian form, i.e. 
\begin{equation}
\mu(d w_1,d w_2) = \nu(w_1, w_2) d w_1 d w_2 = \pi^{-2}\exp{(-|w_1|^2-|w_2|^2}) d w_1 d w_2,
\label{eq:20230621wz01}
\end{equation}
where $\nu(w_1, w_2)$ denotes the measure density.  
The bipartite entangled squeezed coherent states can be represented in the Bargmann space by the normalized wave functions, $\widetilde{\psi}_{k, \rm{coh}}(w_1,w_2;z_1,z_2, \xi)$, which satisfy the resolution of the identity with respect to the Lebesgue measure in the form~\cite{lumini,FHSbook}
\begin{equation}
\frac{1}{\pi^2}\int_{\mathbb{C}^2}{dz_1dz_2}\;
\overline{\widetilde{\psi}_{k, \rm{coh}}(w_1^{\prime},w_2^{\prime};z_1,z_2, \xi)}
\,\widetilde{\psi}_{k, \rm{coh}}(w_1,w_2;z_1,z_2, \xi)
=
K(\bar{w}_1, \bar{w}_2, w^{\prime}_1, w^{\prime}_2),
\label{eq:20231111b}
\end{equation}
where $K(\bar{w}_1, \bar{w}_2, w^{\prime}_1, w^{\prime}_2)$ is the reproducing kernel in the Bargmann space. 
Upon substituting $w^{\prime}_1=w_1$ and $w^{\prime}_2=w_2$ in the reproducing kernel, we obtain the desired measure density, $\nu(w_1, w_2)$~\cite{FHSbook}. 
For determining the form of the wave functions corresponding to the bipartite entangled squeezed coherent states, one can use their expansion in the bipartite orthonormal basis, $|m,n\rangle=|m\rangle\otimes |{n}\rangle$, according to the formula
\begin{equation}
|\widetilde{\psi}_{k,\rm{coh}}(1,2;z_1,z_2, \xi)\rangle
=
\sum_{m,n=0}^{\infty}\varphi_{k,(m,n)}(1,2;z_1,z_2,\xi)
|m,n\rangle,
\label{eq:20230516w02}
\end{equation}
where $\varphi_{k,(m,n)}(1,2;z_1,z_2,\xi)$ are complex coefficients.
Hence, the wave functions of these states can be written in the following form
\begin{equation}
\widetilde{\psi}_{k, \rm{coh}}(w_1,w_2;z_1,z_2, \xi)
=
\sum_{m,n=0}^{\infty}\varphi_{k,(m,n)}(1,2;z_1,z_2,\xi)
\langle w_1,w_2|m,n\rangle,
\label{eq:20230516w03}
\end{equation} 
If elements of basis $\langle w_1,w_2|m,n\rangle$ in the Bargmann space are given by monomials in the form $(m! n!)^{-1/2}\overline{w}_1^{\,m}\overline{w}_2^{\,n}$, then the coefficients $\varphi_{k,(m,n)}(1,2;z_1,z_2,\xi)$ can be expressed by the holomorfic Hermite function in two variables. 
According to results presented in Refs.~\cite{lumini,Gorska2018} these functions are denoted by $\varphi_{k,(m,n)}^{(\alpha)}(z_1,z_2)$  with  the parameter $\alpha\in(0,1)$ which is related to the squeezing parameter $\xi$ through formula
\begin{equation}
\xi
=
\textrm{artanh}\left({\frac{1-\alpha}{1+\alpha}}\right)=-\frac{1}{2}\ln{\alpha}.
\label{eq:20221014}    
\end{equation}
In what follows the arguments $1$ and $2$ numbering the modes are omitted which simplified our previous notation.
Bearing in mind that the squeezing of the base states can be done in twofold ways, we can express the complex functions $\varphi_{k, (m,n)}^{(\alpha)}(z_1,z_2)$ either in product or nonproduct form. 
The product form reads
\begin{equation}
\varphi_{1, (m,n)}^{(\alpha)}(z_1,z_2) 
= 
h_m^{(\alpha)}(z_1)h_n^{(\alpha)}(z_2),
\quad
m, n = 0, 1, 2 \ldots
\label{eq:20220722w12}    
\end{equation}
where $h_n^{(\alpha)}(z)$ stands for the holomorfic Hermite functions of complex variable $z$,
\begin{equation}
h_{n}^{(\alpha)}(z) 
= 
\left(\frac{2\sqrt{\alpha}}{1+\alpha}\right)^{\!\frac{1}{2}} \left(\frac{1-\alpha}{1+\alpha}\right)^{\!\frac{n}{2}} 
\frac{e^{\frac{1-\alpha}{1+\alpha}
\frac{z^{2}}{2}}}{\sqrt{2^{n} n!}} 
H_{n}\left(\sqrt{\frac{2\alpha}{1-\alpha^{2}}}z\right).
\label{eq:20220722w13}
\end{equation}
The symbol $H_n(\cdot)$ in this formula corresponds to the holomorphic Hermite polynomials of order~$n$, which the general form is following~\cite{SLLvanEijndhoven90},
\begin{equation}
H_{n}(z)
=n!\sum_{m=0}^{\infty}\frac{(-1)^{m}(2z)^{n-2m}}{m!(n-2m)!}, 
\label{eq:20220820w1}
\end{equation}
and the orthogonality relation reads~\cite{SLLvanEijndhoven90}
\begin{equation}
\int_{\mathbb{C}} H_m(z) \overline{H_n(z)} e^{-(1-\alpha)x^2 - (\frac{1}{\alpha} - 1)y^2} dx dy 
= 
\frac{\pi\sqrt{\alpha}}{1-\alpha} \left(2\frac{1+\alpha}{1-\alpha}\right)^n n! \delta_{mn}.
\label{eq:20220820w2}
\end{equation}
It is worth noting that the polynomials given by Eq.~(\ref{eq:20220820w1}) were introduced in Ref.~\cite{SLLvanEijndhoven90} for the first time, and according to the inventors bear their name, i.e. they are also known as the Eijndhoven-Meyers polynomials~\cite{SLLvanEijndhoven90}.
Owing to the orthogonality relation given by Eq.~(\ref{eq:20220820w2}) can be shown that the considered holomorphic Hermite polynomials $h_{m}^{(\alpha)}(z)$ form an orthonormal set with respect to the Gaussian measure \cite{STAli2014}, likewise as in the case of the harmonic oscillator coherent states.

The second class of analytic functions, $\varphi_{2, (m,n)}^{(\alpha)}(z_1,z_2)$, to be considered by us is generated by the holomorphic Hermite functions of the two complex variables, $h_{m,n}^{(\alpha)}(z_1,z_2)$,  with $m, n=0,1,\ldots$, and $0 < \alpha < 1$, i.e. we assume that
\begin{equation}
\varphi_{2, (m,n)}^{(\alpha)}(z_1,z_2)
=
h_{m,n}^{(\alpha)}(z_1,z_2). 
\label{eq:20220722w14-00}    
\end{equation}
These holomorphic Hermite functions are expressed as follows~\cite{Gorska2018}
\begin{equation}
h_{m, n}^{(\alpha)}(z_{1}, z_{2}) 
= 
\frac{2\sqrt{\alpha}}{1+\alpha} 
\left(\frac{1-\alpha}{1+\alpha}\right)^{\!\frac{m+n}{2}} 
\frac{e^{\frac{1-\alpha}{1+\alpha} z_{1} z_{2}}}{\sqrt{m! n!}} 
H_{m, n}\left(\frac{2\sqrt{\alpha}}{\sqrt{1-\alpha^{2}}} z_{1}, \frac{2\sqrt{\alpha}}{\sqrt{1-\alpha^{2}}} z_{2}\right), 
\label{eq:20220722w14}
\end{equation}
where $z_1, z_2 \in \mathbb{C}$ and the symbol $H_{m, n}(\cdot,\cdot)$ denotes the complex Hermite polynomials defined by the formula~\cite{Gorska2018} 
\begin{equation}
H_{m, n}(z_{1}, z_{2}) 
= 
\sum_{k=0}^{\min\{m, n\}} 
\binom{m}{k}\binom{n}{k} 
(-1)^k k! z_{1}^{m-k} z_{2}^{n-k}
\label{eq:20220820w6}
\end{equation}
whose orthogonality is treated by \cite[Theorem 3]{Gorska2018}. 
We emphasize that the Hermite functions~(\ref{eq:20220722w14}) form the orthonormal basis with respect to the Gaussian measure~(\ref{eq:20230621wz01}).
The algebraic and analytic properties of these polynomials are considered in Refs.~\cite{Gorska2018, MEHIsmail16,AWunsche15,ghanmi1,ghanmi2}. 
In particular, it is also worth noting that the complex Hermite polynomials, $H_{m, n}(z_{1}, z_{2})$ for $z_{2}=\bar{z}_1$  reduce themselves to the polynomials of the real variables $x,y$ with complex coefficient~\cite{STAli2014, KIto52, NCotfas10}. 
For the first time, this kind of polynomials, i.e. $H_{m, n}(z_{1}, {\bar z_{1}})$ appeared in Ito's work~\cite{KIto52} on applications of complex random variables in stochastic processes. 
Hence, they are called the Ito polynomials. 
Let us note that the polynomials $H_{m, n}(z_{1}, {\bar z_{1}})$ were also introduced and discussed in the physical context~\cite{Fan_PRA49p704y1994}.
Finally,  note that both the functions given by Eq.~(\ref{eq:20220722w12}) and Eq.(\ref{eq:20220722w14-00}) satisfy the condition in the form 
\begin{equation}
\sum_{m,n=0}|\varphi_{k, (m,n)}^{(\alpha)}(z_1,z_2)|^2
<
\infty, 
\quad\textrm{for all $z_1, z_2\in\mathbb{C}$},
\label{eq:20231111a}    
\end{equation}
which guarantees normalizability of the wave functions $\widetilde{\psi}_{k, \rm{coh}}(w_1,w_2;z_1,z_2, \xi)$. 

The bipartite entangled squeezed coherent states can be represented in the Hilbert space of the square-integrable functions ${\mathcal{L}^{2}}(\mathbb{R}^2, dx_1dx_2)$ by Schr{\"o}dinger's wave functions $\widetilde{\psi}_{k, \rm{coh}}(x_1,x_2;z_1,z_2, \xi)$ for the index $k=1,2$. 
To determine these functions, we write down Eq.~(\ref{eq:20230516w02}) in the Schr{\"o}dinger representation, i.e.
\begin{equation}
\widetilde{\psi}_{k,\rm{coh}}^{(\alpha)}(x_1,x_2; z_1,z_2)
:=
\langle x_1, x_2|\widetilde{\psi}_{k,\rm{coh}}^{(\alpha)}(z_1,z_2)\rangle
=
\sum_{m,n=0}^{\infty}\varphi_{k, (m,n)}^{(\alpha)}(z_1,z_2)
\langle x_1, x_2|m,n\rangle,
\label{eq:20231103w01}
\end{equation}
where the explicit form of the functions $\langle x_1, x_2|m,n\rangle$ has the form
\begin{equation}
\langle x_1, x_2|m,n\rangle
=
\sqrt{\frac{ab}{\pi}\frac{1}{2^{m+n}m!n!}}
\exp{\left[
-\frac{1}{2}
\left(a^2x_1^2+b^2x_2^2\right)
\right]}
H_m(ax_1)H_n(bx_2),
\label{eq:20231103w02}
\end{equation}
where $H_m(\cdot)$ denotes the standard Hermite polynomial of real variable, and $a$, $b$ are positive parameters  correspond to the inverse oscillator lengths along $x_1$ and $x_2$ directions, respectively. 
As mentioned above, the form of the function $\varphi_{k, (m,n)}^{(\alpha)}(z_1,z_2) $ explicitly depends on the choice of the index $k$. 
In the case $k=1$, this function is given by Eq.~(\ref{eq:20220722w12}), and direct application of the Mehler formula~\cite{AWunsche15} to the RHS of Eq.~(\ref{eq:20231103w01}) leads us to the following normalized Schr{\"o}dinger's wave function
\begin{eqnarray}
\widetilde{\psi}_{1,\rm{coh}}^{(\alpha)}(x_1,x_2; z_1,z_2)
&=&
\sqrt{\frac{ab}{\pi\alpha}} 
\exp\left[
-\frac{a^2}{2\alpha}\left(x_1-\frac{\sqrt{2\alpha}}{a}z_{1,r}\right)^2
-\frac{b^2}{2\alpha}\left(x_2-\frac{\sqrt{2\alpha}}{b}z_{2,r}\right)^2\right.
\nonumber\\
& + &
\left. 
i\sqrt{\frac{2}{\alpha}}\left(a x_1z_{1,i}+bx_2z_{2,i}\right)
-i\left(z_{1,r}z_{1,i}+z_{2,r}z_{2,i}\right)
\right],
\label{eq:20231103w04}
\end{eqnarray}
where $z_{q,r}$ and $z_{q,i}$ denote the real and imaginary part of the complex number $z_q$, respectively for $q=1, 2$.
In turn, for the case  $k=2$, the function $\varphi_{2, (m,n)}^{(\alpha)}(z_1,z_2) $ is given by Eq.~(\ref{eq:20220722w14-00}). 
Applying a similar procedure to the RHS of Eq.~(\ref{eq:20231103w01}), we obtain the normalized Schr{\"o}dinger's wave function in the form
\begin{eqnarray}
\widetilde{\psi}_{2,\rm{coh}}^{(\alpha)}(x_1,x_2; z_1,z_2)
&=&
\sqrt{\frac{ab}{\pi}}
\exp\left[
-\frac{1+\alpha^2}{4\alpha}a^2\left(x_1-\frac{(\alpha+1)z_{1,r}+(\alpha-1)z_{2,r}}{a\sqrt{2\alpha}}\right)^2\right.
\nonumber\\
&-&
\left. 
\frac{1+\alpha^2}{4\alpha}b^2\left(x_2-\frac{(\alpha-1)z_{1,r}+(\alpha+1)z_{2,r}}{b\sqrt{2\alpha}}\right)^2\right]
\nonumber\\
&\times&
\exp\left[
-\frac{1-\alpha^2}{2\alpha}ab\left(x_1-\frac{(\alpha+1)z_{1,r}+(\alpha-1)z_{2,r}}{a\sqrt{2\alpha}}\right)
\right.
\nonumber\\
&\times&
\left(x_2-\frac{(\alpha-1)z_{1,r}+(\alpha+1)z_{2,r}}{b\sqrt{2\alpha}}\right)
\left.
-
i z_{1,r} z_{1,i}-i z_{2,r} z_{2,i}\right]
\nonumber\\
&\times&
\exp\left[i\frac{ax_1}{\sqrt{2\alpha}}\left((1+\alpha)z_{1,i}+(1-\alpha)z_{2,i}\right)\right.
\nonumber\\
&+&
\left.
i\frac{bx_2}{\sqrt{2\alpha}}\left((1+\alpha)z_{2,i}+(1-\alpha)z_{1,i}\right)\right].
\label{eq:20231103w06}
\end{eqnarray}

Alternatively, Schr{\"o}dinger's wave functions expressed by Eqs~(\ref{eq:20231103w04}) and (\ref{eq:20231103w06}) can be derived by applying the inverse of the Segal-Bargmann transform~\cite{VBargmann61}, which allows one to convert the Bargmann representation of the wave function into its Schr{\"o}dinger's representation according to the formula
\begin{equation}
\widetilde{\psi}_{k, \rm{coh}}(x_1,x_2;z_1,z_2, \xi)
\frac{1}{\pi^2}
\int{dw_1dw_2}\;
\langle x_1,x_2|w_1,w_2\rangle
\widetilde{\psi}_{k, \rm{coh}}(w_1,w_2;z_1,z_2, \xi),
\label{eq:20230516w01}    
\end{equation}
where $\langle x_1,x_2|w_1,w_2\rangle$ represents the integral kernel of the Segal-Bargmann transform and is given by the formula~\cite{VBargmann61}
\begin{eqnarray}
\langle x_1, x_2| w_{1}, w_{2}\rangle 
&=&
\sqrt{\frac{a b}{\pi}} 
\exp{\left[
-\frac{1}{2}(w_{1}^{2} 
+ 
w_{2}^{2} + a^{2} x_1^{2} + b^{2} x_2^{2})
\right]}
\nonumber\\
&\times &
\exp{\left[\sqrt{2}(ax_1 w_{1}  + bx_2 w_{2} )-|w_1|^2-|w_2|^2\right]}.
\label{eq:20230530w01}
\end{eqnarray}
However, this approach is technically more tedious.
%%%%%%%%%%%%%%%%%%%%%%%%%%%%%%%%%%%%%%%%%%%%%%%%%%%%%%%%%%%%%%%%%%%%%%%%%%%%%%%%%%%%%%%%
\section{\label{sec:3} Results and discussion}
%%%%%%%%%%%%%%%%%%%%%%%%%%%%%%%%%%%%%%%%%%%%%%%%%%%%%%%%%%%%%%%%%%%%%%%%%%%%%%%%%%%%%%%%

%%%%%%%%%%%%%%%%%%%%%%%%%%%%%%%%%%%%%%%%%%%%%%%%%%%%%%%%%%%%%%%%%%%%%%%%%%%%%%%%%%%%%%%%
\subsection{\label{ssec:WDF}Wigner representation}
%%%%%%%%%%%%%%%%%%%%%%%%%%%%%%%%%%%%%%%%%%%%%%%%%%%%%%%%%%%%%%%%%%%%%%%%%%%%%%%%%%%%%%%%
The phase-space representation is often used to examine physical systems, such as electromagnetic modes in optical or microwave cavities, or mechanical modes of harmonic oscillators, where degrees of freedom constitute continuous variables. 
This approach, frequently applied in quantum optics, is intuitive and robust, facilitating the investigation of the system non-classical state features or their dynamics. 
According to this approach, the system states within the phase space are represented by non-classical distribution functions of conjugate variables defined on a real symplectic space $\mathbb{R}^2\times\mathbb{R}^2$ equipped with canonical symplectic form, relevant to these variables. 
Among numerous such functions, the Wigner distribution function (WDF) plays a pivotal role, providing a lucid phase-space representation of the quantum state and emphasising the cognitive aspect of the obtained results. 
Let us first start by introducing the WDF for a bipartite pure state $\left| f \right>$ represented by the Schr{\"o}dinger wave function $f(x_1,x_2)\in{\mathcal{L}^{2}}(\mathbb{R}^2, dx_1dx_2)$ according to the definition~\cite{Wigner_PR40p749y1932,Gosson2017} 
\begin{eqnarray}
W\left(f\right)(x_1,x_2,p_1,p_2)
&=&
\frac{1}{(2\pi\hbar)^2}
\int_{\mathbb{R}^2}\!\!{dX_1dX_2}
\exp\left[-\frac{i}{\hbar}\left(p_1X_1+p_2X_2\right)
\right]
\nonumber\\
&\times&
f\left(x_1+\frac{X_1}{2},x_2+\frac{X_2}{2}\right)
\overline{f\left(x_1-\frac{X_1}{2},x_2-\frac{X_2}{2}\right)},
\label{eq:20231120-a}    
\end{eqnarray}
where $p_1$ and $p_2$ are momentum variables, and $x_1$ and $x_2$ are position variables.
The transfer of our original problem to the phase-space representation and the application of the WDF is partially motivated by the form of the Schr{\"o}dinger wave functions of bipartite states given by Eqs~(\ref{eq:20231103w04}) and (\ref{eq:20231103w06}).
Let us note that both these wave functions have the form of the Gaussian functions shifted to different points of the configurational space and with different phase factors. 
This observation allows us to apply the following de Gosson result for the generalized Gaussian functions of the Wigner distribution function~\cite[Corollary 113]{Gosson2017} adapted to our needs, namely
\begin{corollary}
Let $\bf{M}$ be a positive definite real symmetric $n\times n$ matrix, and ${\bf{r}}^T$ be a $1\times n$ row matrix of the configurational variables $(x_1, x_2,\ldots, x_n)$.
The Wigner transform of the normalized generalized Gaussian
\begin{equation}
\psi_{\bf{M}}({\bf{r}})
=
\left(\frac{1}{\pi}\right)^{n/4}
\left(\det{\bf{M}}\right)^{1/4}
\exp{\left(-\frac{1}{2}{\bf{r}}^T{\bf{M}}{\bf{r}}\right)}
\label{eq:20231202-a}
\end{equation}
is the phase-space Gaussian
\begin{equation}
W(\psi_{\bf{M}})(\boldsymbol{\gamma})
=
\left(\frac{1}{\pi\hbar}\right)^n
\exp{\left(-\frac{1}{2}{\boldsymbol{\gamma}}^T{\boldsymbol{\Sigma}^{-1}}{\boldsymbol{\gamma}}\right)},    
\label{eq:20231202-b}
\end{equation}
where ${{\boldsymbol{\Sigma}}}$ is the covariance matrix in the form
\begin{equation}
{\boldsymbol{\Sigma}}
=
\left[
\begin{array}{ccc}
\frac{1}{2}{\bf{M}}^{-1} & \bf{0} \\
\bf{0} & \frac{\hbar^2}{2}\bf{M} 
\end{array}
\right],
\label{eq:20231202-c}
\end{equation}
and $\boldsymbol{\gamma}^T$ is a $1\times n$ row matrix of the phase-space variables $(x_1,x_2,\ldots,x_n, p_1, p_2,\ldots, p_n)$.
\label{col:20231202-d}
\end{corollary}
However, the direct application of the quoted corollary requires centring the Gaussians given by Eqs~(\ref{eq:20231103w04}) and (\ref{eq:20231103w06}) at the origin. 
We obtain this result by utilizing the Heisenberg-Weyl operator to the aforementioned Schr{\"o}dinger wave functions. 
Let us note that these wave functions can be regarded as a result of the acting of the Heisenberg-Weyl operator, $\widehat{T}({y}_{1,k}, y_{2,k}, q_{1,k}, q_{2,k})$, on the unshifted Gaussian functions, $\widetilde{\psi}_{k,\rm{coh}}^{(\alpha)}(x_1,x_2; 0,0)$, that is,
\begin{equation}
\widetilde{\psi}_{k,\rm{coh}}^{(\alpha)}(x_1,x_2; z_1,z_2)
=
\widehat{T}({y}_{1,k}, y_{2,k}, q_{1,k}, q_{2,k})\widetilde{\psi}_{k,\rm{coh}}^{(\alpha)}(x_1,x_2; 0,0),
\label{eq:HW-operator-2}  
\end{equation}
where $y_{j,k}$ and $q_{j,k}$ are the real and imaginary parts of complex numbers $z_{j}$, respectively, for $j=1,2$ and $k$ fixed.
According to Definition 2 in Ref.~\cite{Gosson2017}, the explicit form of the Heisenberg-Weyl operator is the following
\begin{eqnarray}
\widehat{T}({y}_{1,k}, y_{2,k}, q_{1,k}, q_{2,k})f(x_1,x_2)
&=&
\exp{\left[
\frac{i}{\hbar}
\left(
q_{1,k}x_1+q_{2,k}x_2\right)
-
\frac{1}{2}q_{1,k}y_{1,k}-\frac{1}{2}q_{2,k}y_{2,k}
\right]}
\nonumber\\
&\times&
f(x_1-y_{1,k}, x_2-y_{2,k}). 
\label{eq:20231118-a}    
\end{eqnarray}
Substituting Eq.~(\ref{eq:20231118-a}) into the definition~(\ref{eq:20231120-a}), we obtain the result
\begin{equation}
W\left(
\widehat{T}({y}_{1,k}, y_{2,k}, q_{1,k}, q_{2,k})f\right)(x_1,x_2,p_1,p_2)
=W\left(f\right)(x_1-{y}_{1,k}, x_2-{y}_{2,k}, p_1-q_{1,k}, p_2-q_{2,k})
\label{eq:20231127-01}    
\end{equation}
which relates the effect of the Heisenberg-Weyl operator acting on the wave function with the Wigner distribution function of this function, and it allows us to determine the unshifted form of the phase-space Gaussian. 
After some algebraic manipulations, we identified the parameters of the Heisenberg-Weyl operator and the wave functions~(\ref{eq:20231103w04}) and (\ref{eq:20231103w06}) separately to each of the indexes $k$. 
As a consequence, we received two distinct results. 
For the index $k=1$, the parameters of the Heisenberg-Weyl operator have the form 
\begin{equation}
y_{1,1}=\frac{\sqrt{2\alpha}}{a}z_{1,r},
\quad
y_{2,1}=\frac{\sqrt{2\alpha}}{b}z_{2,r},
\quad
q_{1,1}=\sqrt{\frac{2}{\alpha}}az_{1,i},
\quad
q_{2,1}=\sqrt{\frac{2}{\alpha}}bz_{2,i},
\label{eq:20231118-b}    
\end{equation}
and the corresponding unshifted wave function, $\widetilde{\psi}_{1,\rm{coh}}^{(\alpha)}(x_1,x_2; 0,0)$, is given by the formula
\begin{equation}
\widetilde{\psi}_{1,\rm{coh}}^{(\alpha)}(x_1,x_2; 0,0)
=
\sqrt{\frac{ab}{\pi\alpha}} 
\exp{\left[-\frac{1}{2}\left(\frac{a^2}{\alpha}x_1^2+\frac{b^2}{\alpha}x_2^2\right)\right]},
\label{eq:20231118-c}
\end{equation}
for which the matrix of the quadratic form, ${\bf{M}}_{1,\alpha}$, in two variables $(x_1, x_2)$  defining this wave function has the diagonal form
\begin{equation}
{\bf{M}}_{1,\alpha} 
=
\frac{1}{\alpha}\textrm{diag}\left(a^2, b^2\right).
\label{eq:20240108-a}    
\end{equation}
In turn, for the index $k=2$, the parameters of the Heisenberg-Weyl operator read as follows
\begin{eqnarray}
y_{1,1} = \frac{(\alpha+1)z_{1,r}+(\alpha-1)z_{2,r}}{a\sqrt{2\alpha}}, 
&\quad&
y_{2,1} = \frac{(\alpha-1)z_{1,r}+(\alpha+1)z_{2,r}}{b\sqrt{2\alpha}},
\nonumber\\
q_{1,1} = \frac{a}{\sqrt{2\alpha}}\left[(1+\alpha)z_{1,i}+(1-\alpha)z_{2,i}\right],
&\quad&
q_{2,1} = \frac{b}{\sqrt{2\alpha}}\left[(1+\alpha)z_{2,i}+(1-\alpha)z_{1,i}\right],
\label{eq:20231118-d}
\end{eqnarray}
and the unshifted wave function, $\widetilde{\psi}_{2,\rm{coh}}^{(\alpha)}(x_1,x_2; 0,0)$, has the form
\begin{equation}
\widetilde{\psi}_{2,\rm{coh}}^{(\alpha)}(x_1,x_2; 0,0)
=
\sqrt{\frac{ab}{\pi}}
\exp{\left[-\frac{1}{2}\left(\frac{1+\alpha^2}{2\alpha}a^2 x_1^2+\frac{1+\alpha^2}{2\alpha}b^2 x_2^2+\frac{1-\alpha^2}{\alpha}abx_1x_2\right)\right]}.
\label{eq:20231118-e}
\end{equation}
The matrix of the quadratic form, ${\bf{M}}_{2,\alpha}$, in two variables $(x_1, x_2)$ associated with the wave function, $\widetilde{\psi}_{2,\rm{coh}}^{(\alpha)}(x_1,x_2; 0,0)$, has the following form
\begin{equation}
{\bf{M}}_{2,\alpha} 
=
\frac{1}{2\alpha}
\left[
\begin{array}{cccc}
(1+\alpha^2)a^2 & (1-\alpha^2)ab \\
(1-\alpha^2)ab & (1+\alpha^2)b^2 
\end{array}
\right].
\label{eq:20240108-b}    
\end{equation}
Let us note that the form of the wave functions given by Eqs~(\ref{eq:20231118-c}) and (\ref{eq:20231118-e}) are consistent with the assumptions of the Corollary~\ref{col:20231202-d}. 
Hence, the Wigner distribution functions corresponding to these wave functions can be directly expressed by Eq.~(\ref{eq:20231202-b}), wherein the relations between matrices ${\bf{M}}_{k,\alpha}$ and ${\bf{\boldsymbol{\Sigma}}}_{k,\alpha}$ are given by Eq.~(\ref{eq:20231202-c}), for $k=1, 2$. 
However, the covariance matrices corresponding to these two cases are different. 
For the Wigner distribution function corresponding to the wave function $\widetilde{\psi}_{1,\rm{coh}}^{(\alpha)}(x_1,x_2; 0,0)$, the covariance matrix has the form
\begin{equation}
{\bf{\boldsymbol{\Sigma}}}_{1,\alpha} 
=
\frac{1}{2}
\textrm{diag}\left(\frac{\alpha}{a^2}, \frac{\alpha}{b^2}, \frac{a^2\hbar^2}{\alpha}, \frac{b^2\hbar^2}{\alpha}\right),
\label{eq:20240108-c}    
\end{equation}
while the covariance matrix associated with the Wigner distribution function corresponding to the wave function $\widetilde{\psi}_{2,\rm{coh}}^{(\alpha)}(x_1,x_2; 0,0)$ is given by the formula 
\begin{equation}
{\bf{\boldsymbol{\Sigma}}}_{2,\alpha} 
=
\frac{1}{4\alpha}
\left[
\begin{array}{cccc}
\frac{1+\alpha^2}{a^2} & \frac{\alpha^2-1}{ab} & 0 & 0\\
\frac{\alpha^2-1}{ab} & \frac{1+\alpha^2}{b^2} & 0 & 0\\
0 & 0 & (1+\alpha^2)a^2\hbar^2 & (1-\alpha^2)ab\hbar^2\\
0 & 0 & (1-\alpha^2)ab\hbar^2 & (1+\alpha^2)b^2\hbar^2
\end{array}
\right].
\label{eq:20240108-d}    
\end{equation}
Each of the obtained covariance matrices, ${\bf{\boldsymbol{\Sigma}}}_{k,\alpha}$ corresponding to Gaussian's Wigner distribution functions in the form given by Eq.~(\ref{eq:20231202-b}), for  $k=1, 2$,  satisfies the Robertson-Schr{\"o}dinger uncertainty relation in the form
\begin{equation}
{\bf{\boldsymbol{\Sigma}}}_{k,\alpha}
+
\frac{i\hbar}{2}{\bf{J}}
\geqslant
0,
\label{eq:20240121-a}
\end{equation}
where ${\bf{J}}$ is the standard symplectic matrix (${\bf{J}}^T=-{\bf{J}}$, ${\bf{J}}^2=-{\bf{I}}$, where ${\bf{I}}$ is the identity matrix). 
Let us note that the appearance of the matrix ${\bf{J}}$ in the aforementioned inequality is a consequence of the following position and momentum operators ordering in the row: $\hat{\boldsymbol{\gamma}}^T=[\hat{x}_1\;\hat{x}_2\;\hat{p}_1\;\hat{p}_2]$ and consequently, the canonical commutation relation takes the form: $\left[\hat{\gamma}_k,\hat{\gamma}_l\right]=i\hbar J_{kl}{\textrm{Id}}_{L^2(\mathbb{R}^2)}$, where $\hat{\gamma}_k$ refers to the components of vector $\hat{\boldsymbol{\gamma}}$, and $J_{kl}$ is the matrix element of the skew-symmetric matrix ${\bf{J}}$. 

Entanglement properties of the bipartite states can be characterized with the positive partial transpose (PPT) criterion of Peres and Horodecki~\cite{Peres_PRL77p1413y1996,Horodecki_PLA223p1y1996}. 
It is based on the operation of partial transposition of the density operator of the bipartite system obtained by the transposition of indices corresponding to one subsystem only. 
On the other hand, Simon~\cite{Simon_PRL84p2726y2000} generalized this criterion to systems with continuous variables and proved that the PPT criterion forms a necessary and sufficient condition for the separability of all bipartite Gaussian states. 
The key idea for this result is based on the observation that the transpose operation of the density operator transcribes the momentum reversal operation in the Wigner distribution function. 
It goes according to the formula: $W(\psi_{\bf{M}})(x_1,x_2,p_1,p_2)\rightarrow W(\psi_{\bf{M}})(x_1,x_2,p_1,-p_2) $, which corresponds to a mirror reflection that  inverts only the momentum variable $p_2$.

According to Simon's result~\cite{Simon_PRL84p2726y2000}, the Horodecki-Peres separability criterion for continuous variables takes the form
\begin{equation}
{\widetilde{\bf{\boldsymbol{\Sigma}}}}_{k,\alpha}
+
\frac{i\hbar}{2}{\bf{J}}
\geqslant
0,
\label{eq:20240121-b}    
\end{equation}
where the matrix ${\widetilde{{\boldsymbol{\Sigma}}}}_{k,\alpha}$ is given by the formula
\begin{equation}
{\widetilde{\bf{\boldsymbol{\Sigma}}}}_{k,\alpha}
=
{\boldsymbol{\Lambda}}
{{\bf{\boldsymbol{\Sigma}}}}_{k,\alpha}
{\boldsymbol{\Lambda}}^{T}.
\label{eq:20240121-c}    
\end{equation}
In this formula, the matrix ${\boldsymbol{\Lambda}}$ has the form: 
${\boldsymbol{\Lambda}}^{T}=\textrm{diag}\left(1, 1, 1, -1\right)$. 
On the other hand, de~Gosson shows that the Simon-Horodecki-Peres separability criterion is equivalent to the following proposition~\cite[Proposition 168]{Gosson_QHA2021}
\begin{equation}
\lambda_{min}
\geqslant
\frac{\hbar}{2},
\label{eq:20240121-d}    
\end{equation}
where $\lambda_{min}$ is the minimal symplectic eigenvalue of the matrices  ${\widetilde{\bf{\boldsymbol{\Sigma}}}_{k,\alpha}}$. 
Based on this proposition, we determine the symplectic eigenvalues of the matrices ${\widetilde{\boldsymbol{\Sigma}}}_{k,\alpha}$ by solving characteristic equations of the form
\begin{equation}
\textrm{det}({\bf{J}}{\widetilde{\boldsymbol{\Sigma}}}_{k,\alpha}-\lambda{\bf{I}})
=0.
\label{20240121-e}    
\end{equation}
However, let us note the eigenvalues of the matrix ${\bf{J}}{\widetilde{\boldsymbol{\Sigma}}}_{k,\alpha}$ are complex numbers in the form $\pm i\lambda^{(j)}_{k,\alpha}$ with $\lambda^{(j)}_{k,\alpha}>0$ , where the upper index $j$ numbers these eigenvalues. 
The positive numbers $\lambda^{(j)}_{k,\alpha}$ are symplectic eigenvalues of the matrix ${\widetilde{\boldsymbol{\Sigma}}}_{k,\alpha}$ and they form the symplectic spectrum of this matrix.
As a result of our calculations, we obtained the following results.
For matrix ${\widetilde{\bf{\boldsymbol{\Sigma}}}_{1,\alpha}}$, we have one distinguishable symplectic eigenvalue of the form $\lambda^{(1)}_{1,\alpha}=\hbar/2$.
In turn, for the matrix ${\widetilde{\bf{\boldsymbol{\Sigma}}}_{2,\alpha}}$ we have two different symplectic eigenvalues, namely $\lambda^{(1)}_{2,\alpha}=\hbar/(2\alpha)$ and $\lambda^{(2)}_{2,\alpha}=\hbar\alpha/2$.
Based on these results, we can conclude that the symplectic eigenvalue $\lambda^{(1)}_{1,\alpha}=\hbar/2$ is equal to the minimal symplectic eigenvalue of the matrix ${\widetilde{\bf{\boldsymbol{\Sigma}}}_{1,\alpha}}$. 
Thus, the corresponding bipartite state has a separable form. 
However, this result was expected because the covariance matrix associated with this state is diagonal. 
On the other hand, both the symplectic eigenvalues of the matrix ${\widetilde{\bf{\boldsymbol{\Sigma}}}_{2,\alpha}}$ depend on the parameter $\alpha\in(0,1)$. 
Hence, the minimal eigenvalue of this matrix is $\lambda^{(min)}_{2,\alpha}=\lambda^{(2)}_{2,\alpha}$, and thereby the Simon-Horodecki-Peres separability criterion is violated. 
The bipartite state corresponding to this case is entangled.
It is worth giving an in-depth look at this result because it depends on the value of the $\alpha$ parameter. 
In particular, its limiting cases are quite interesting. 
Namely, in the limit, $\alpha\rightarrow 1_{-}$, the covariance matrix ${\bf{\boldsymbol{\Sigma}}}_{2,\alpha}$ is diagonal. 
It means that the corresponding bipartite state becomes disentangled. 
Moreover, it is also worth noting that in the limit $\alpha\rightarrow 1_{-}$, the holomorphic Hermite functions given by Eq.~(\ref{eq:20220722w14}) are simplified to the product form of the monomials. 
In turn, in the limit $\alpha\rightarrow 0_{+}$, the covariance matrix, ${\bf{\boldsymbol{\Sigma}}}_{2,\alpha}$, corresponding to the aforementioned bipartite state remains non-diagonal, meaning the state is still entangled. 
However, the influence of this limit on the holomorphic Hermite functions~(\ref{eq:20220722w14}) was examined only for the case $z_2=\bar{z}_1$, and it was shown that these functions are reduced to the form of the Ito polynomials~\cite{Fan_PRA49p704y1994}.  

The application of the Simon-Horodecki-Peres criterion to the considered bipartite states enables us to identify one them as entangled, i.e. this one characterized by the covariance matrix ${\bf{\boldsymbol{\Sigma}}}_{2,\alpha}$. 
Unfortunately, it does not provide us a measure of their entanglement degree. 
To address this issue, we employ the logarithmic negativity measure~\cite{Vidal_PRA65p032314y2002,Plenio_PRL95p090503y2005} in the form adapted to the bipartite Gaussian state, namely~\cite{Kao_SR8p7394y2018}
\begin{equation}
E_{\mathcal{L}}
=
{\textrm{max}}\left(\log{\frac{\hbar}{2\lambda^{(min)}_{2,\alpha}}},0\right).
\label{eq:20240614-01}    
\end{equation}
It allows to quantify the degree of the entanglement in composite systems.
In the case that we consider the minimal symplectic eigenvalue is $\lambda^{(min)}_{2,\alpha}=\lambda^{(2)}_{2,\alpha}$. 
Thus we obtain the logarithmic negativity in the form
\begin{equation}
E_{\mathcal{L}}
=
{\textrm{max}}(-\log{\alpha},0).
\label{eq:20240705-01}    
\end{equation}
Let us note that for $\alpha\rightarrow 1_{-}$, the considered bipartite state becomes separable with respect to the measure given by Eq.~(\ref{eq:20240705-01}). 
Similarly, with respect to the same measure, in the limit $\alpha\rightarrow 0_{+}$, this state becomes maximally entangled. 
Note also that since the logarithmic negativity measure given by Eq.~(\ref{eq:20240705-01}),  is a decreasing function of the parameter $\alpha$, then we conclude that all intermediate bipartite states, for $\alpha<1$, are entangled states, and as the parameter $\alpha$ increases, the degree of entanglement of that state decreases.

%%%%%%%%%%%%%%%%%%%%%%%%%%%%%%%%%%%%%%%%%%%%%%%%%%%%%%%%%%%%%%%%%%%%%%%%%%%%%%%%%%%%%%%%
\subsection{\label{ssec:ho}Physical model realization}
%%%%%%%%%%%%%%%%%%%%%%%%%%%%%%%%%%%%%%%%%%%%%%%%%%%%%%%%%%%%%%%%%%%%%%%%%%%%%%%%%%%%%%%%
Let us note that the Schr{\"o}dinger wave functions we derived can be regarded as the eigenstates of some two-dimensional physical system.
In light of this observation, we can reconstruct the Hamiltonian model of the two-dimensional quantum-mechanical system, in which the physical realization allows us to verify the results previously presented.
Based on the results so far, we state that in the limit $\alpha\rightarrow 1_{-}$, the considered wave functions reduce to the product form of the Gaussian function, each of which minimizes the uncertainty principle.
Furthermore, these Gaussian functions can be viewed as the ground state of a two-dimensional anisotropic harmonic oscillator. 
This suggests that the sought-after quantum-mechanical system can actually be realized by a model of an anisotropic harmonic oscillator in the form 
\begin{equation}
\hat{H}_{SHO}(1,2)
=
\frac{\hat{p}_x^2}{2M}+\frac{\hat{p}_y^2}{2M} 
+
\frac{1}{2}M\omega_1\hat{x}_1^2+\frac{1}{2}M\omega_2\hat{x}_2^2,
\label{eq:20240301a}   
\end{equation}
where $\omega_1$ and $\omega_2$ are the characteristic frequencies of the simple harmonic oscillator along the $x_1$- and $x_2$-direction, respectively. 
Simultaneously, let us recall (cf. the discussion in Sec.\ref{sec:1}) that the squeezing of bipartite coherent states can be obtained by acting the bipartite squeezing operators $\hat{S}_k(1,2;\xi)$ and the translation operators $\hat{D}(1,2;z_1,z_2)$ on the tensor product of two single-mode vacuum states. 
In consequence, we can expect that the eigenequation corresponding to the Hamiltonian (\ref{eq:20240301a}), i.e.
\begin{equation}
\hat{H}_{SHO}(1,2)\left[|\phi_0(1)\rangle\otimes|\phi_0(2)\rangle\right]
=
\frac{1}{2}\hbar(\omega_1+\omega_2)\left[|\phi_0(1)\rangle\otimes|\phi_0(2)\rangle\right],
\label{eq:20240301d}    
\end{equation}
subjected to the unitary transformation generated by the operator $\hat{D}(1,2;z_1,z_2)\hat{S}_k(1,2;\xi)$ takes the form 
\begin{equation}
\hat{H}_k(1,2;z_1,z_2, \xi)
|\widetilde{\psi}_{k,\rm{coh}}(1,2;z_1,z_2, \xi)\rangle
=
\frac{1}{2}\hbar(\omega_1+\omega_2)|\widetilde{\psi}_{k,\rm{coh}}(1,2;z_1,z_2, \xi)\rangle
\label{eq:20240301b}
\end{equation}
where the Hamiltonian $\hat{H}_k(1,2;z_1,z_2, \xi)$ is given by the formula
\begin{equation}
\hat{H}_k(1,2;z_1,z_2, \xi)
=
\hat{D}(1,2;z_1,z_2)\hat{S}_k(1,2;\xi)\hat{H}_{SHO}(1,2)\hat{S}_k^{\dagger}(1,2;\xi)\hat{D}^{\dagger}(1,2;z_1,z_2).
\label{eq:20240301c}    
\end{equation}
The Hamiltonians  $\hat{H}_k(1,2;z_1,z_2, \xi)$ obtained in this manner still describe systems of the anisotropic harmonic oscillators, albeit with the requisite modifications. 
Therefore, it seems reasonable to presume that the construction of these Hamiltonians is based on the assumption that there exists a pair of the oscillator annihilation operators $\hat{c}_1$ and $\hat{c}_2$, and their conjugate creation operators $\hat{c}^{\dagger}_1$ and $\hat{c}^{\dagger}_2$, which are elements of an algebra based on the canonical commutation relations: $[\hat{c}_i,\hat{c}_j^{\dagger}]=\hat{1}\delta_{ij}$, with $[\hat{c}_i,\hat{c}_j]=\hat{0}$, where $i,j=1,2$. 
According to this statement, we can assume that the annihilation and the creation operators are given by 
\begin{eqnarray}
\hat{c}_{1}^{\dag}|m, n\rangle  =  \sqrt{m+1}|m+1, n\rangle \quad &\textrm{and}& \quad \hat{c}_{1}|m, n\rangle = \sqrt{m}|m-1, n\rangle \label{eq:20220722w24}\\ \hat{c}_{2}^{\dag}|m, n\rangle = \sqrt{n+1}|m, n+1\rangle \quad &\textrm{and}& \quad \hat{c}_{2}|m, n\rangle = \sqrt{n}|m, n-1\rangle.
\label{eq:20240301e}
\end{eqnarray}
By using these operators, the occupation number representation of the Hamiltonian of the anisotropic harmonic oscillator~(\ref{eq:20240301a}) can be expressed in the following form
\begin{equation}
\hat{H}_{SHO}(1,2)
=
\hbar\omega_1\hat{c}_{1}^{\dag}\hat{c}_{1}+\hbar\omega_2\hat{c}_{2}^{\dag}\hat{c}_{2}
+
\frac{1}{2}\hbar\left(\omega_1+\omega_2\right).
\label{eq:20240322-1}    
\end{equation}
Subsequently, by applying the transformation given by the formula~(\ref{eq:20240301c} to the Hamiltonian~(\ref{eq:20240322-1}), we obtain the new form of the Hamiltonian, namely
\begin{equation}
\hat{H}_k(1,2; z_1,z_2, \xi)
=
\hbar\omega_1\hat{C}_{k; 1}^{\dag}\hat{C}_{k; 1}
+
\hbar\omega_2\hat{C}_{k; 2}^{\dag}\hat{C}_{k; 2}
+
\frac{1}{2}\hbar\left(\omega_1+\omega_2\right),
\label{eq:20240322-2}    
\end{equation}
in which the new pairs of the creation operators $\hat{C}^{\dagger}_{k; i}$ and the annihilation operators $\hat{C}_{k; i}$, for $i=1,2$ are given by the formulae
\begin{eqnarray}
\hat{C}_{k; 1}^{\dag}  &=&  \hat{D}(1,2;z_1,z_2)\hat{S}_k(1,2;\xi)\hat{c}_{1}^{\dagger}\hat{S}_k^{\dagger}(1,2;\xi)\hat{D}^{\dagger}(1,2;z_1,z_2),
\label{eq:20240322-3}
\\
\hat{C}_{k; 1}  &=&  \hat{D}(1,2;z_1,z_2)\hat{S}_k(1,2;\xi)\hat{c}_{1}\hat{S}_k^{\dagger}(1,2;\xi)\hat{D}^{\dagger}(1,2;z_1,z_2), 
\label{eq:20240322-4}
\end{eqnarray}
and
\begin{eqnarray}
\hat{C}_{k; 2}^{\dag}  &=&  
\hat{D}(1,2;z_1,z_2)
\hat{S}_k(1,2;\xi)\hat{c}_{2}^{\dag}\hat{S}_k^{\dagger}(1,2;\xi)\hat{D}^{\dagger}(1,2;z_1,z_2), 
\label{eq:eq:20240322-5}
\\
\hat{C}_{k; 2}  &=&  \hat{D}(1,2;z_1,z_2)\hat{S}_k(1,2;\xi)\hat{c}_{2}\hat{S}_k^{\dagger}(1,2;\xi)\hat{D}^{\dagger}(1,2;z_1,z_2), 
\label{eq:20240322-6}
\end{eqnarray}
where they satisfy the new cannonical commutation relations in the form: $[\hat{C}_{k; i},\hat{C}_{k; j}^{\dagger}]= (\hat{1}\delta_{ij})_k$, $[\hat{C}_{k; i},\hat{C}_{k; j}]=\hat{0}$, $[\hat{C}_{k; i}^{\dagger},\hat{C}_{k; j}^{\dagger}]= \hat{0}$,
for $i,j=1,2$ and fixed $k$.
In general, each of the operators expressed by the formulas (\ref{eq:20240322-3})--(\ref{eq:20240322-6}) can be transformed into the form
\begin{eqnarray}
\hat{C}_{k; i}^{\dag}  
=  
\sum_{j=1}^2
\left[\mu_{k;ij}\hat{c_j}^{\dagger}+\widetilde{\mu}_{k;ij}\hat{c_j}\right]
+
\xi_{k;i},
\quad \textrm{and} \quad
\hat{C}_{k; i}  
=  
\sum_{j=1}^2
\left[ \nu_{k;ij}\hat{c_j}^{\dagger}+\widetilde{\nu}_{k;ij}\hat{c_j}\right]+\zeta_{k;i}, 
\label{eq:20240322-7}
\end{eqnarray}
where the coefficients $\mu_{k;ij}$ and $\nu_{k;ij}$ are determined for the index $k=2$ and for $i=1, 2$ in the further part of this section. 
The case of $k=1$ has been considered many times in other works~\cite{lumini,STAli2014}, so we do not address it here. 
\begin{equation}
\mu_{2;11}
=
\frac{1+\alpha}{2\sqrt{\alpha}},
\quad
\mu_{2; 12}
=
0,
\quad
{\mu}_{2;21}
=
0,
\quad
\mu_{2;22}
=
\frac{1+\alpha}{2\sqrt{\alpha}},
\label{eq:20240503-1}    
\end{equation}
\begin{equation}
\widetilde{\mu}_{2;11}
=
0,
\quad
\widetilde{\mu}_{2;12}
=
-\frac{1-\alpha}{2\sqrt{\alpha}},
\quad
\widetilde{\mu}_{2;21}
=
-\frac{1-\alpha}{2\sqrt{\alpha}},
\quad
\widetilde{\mu}_{2;22}
=
0,
\label{eq:20240503-2}    
\end{equation}
\begin{equation}
\nu_{2;11}
=
-\frac{1-\alpha}{2\sqrt{\alpha}},
\quad
\nu_{2;12}
=
0,
\quad
\nu_{2;21}
=
-\frac{1-\alpha}{2\sqrt{\alpha}},
\quad
{\nu}_{2;22}
=
0,
\label{eq:20240503-3}    
\end{equation}
\begin{equation}
\widetilde{\nu}_{2;11}
=0,
\quad
\widetilde{\nu}_{2;12}
=
\frac{1+\alpha}{2\sqrt{\alpha}},
\quad
\widetilde{\nu}_{2;21}
=0,
\quad
\widetilde{\nu}_{2;22}
=
\frac{1+\alpha}{2\sqrt{\alpha}}.
\label{eq:20240503-4}    
\end{equation}
In turn, the shifts $\xi_{k; i}$ and $\zeta_{k; i}$ are, respectively
\begin{equation}
\xi_{2; 1} 
= 
\frac{1+\alpha}{2\sqrt{\alpha}}\bar{z}_1 - \frac{1-\alpha}{2\sqrt{\alpha}}z_2 , 
\quad 
\zeta_{2; 1} 
= 
\frac{1+\alpha}{2\sqrt{\alpha}}z_1 - \frac{1-\alpha}{2\sqrt{\alpha}}\bar{z}_2,
\label{eq:20240419-01}  
\end{equation}
\begin{equation}    
\xi_{2; 2} 
=
\frac{1+\alpha}{2\sqrt{\alpha}}\bar{z}_2 - \frac{1-\alpha}{2\sqrt{\alpha}}z_1, 
\quad 
\zeta_{2; 2} 
= 
\frac{1+\alpha}{2\sqrt{\alpha}}z_2 - \frac{1-\alpha}{2\sqrt{\alpha}}\bar{z}_1.
\label{eq:20240419-02}    
\end{equation}
After using the results of (\ref{eq:20240503-1}) -- (\ref{eq:20240419-02}), we can express, in an explicit form, the creation operators $\hat{C}_{2; i}^{\dag}$ and annihilation operators $\hat{C}_{2; i}$ expressed by the formula (\ref{eq:20240322-7}), for $i=1,2$, as follows
\begin{equation}
\hat{C}^\dagger_{2; 1} 
= 
\frac{1+\alpha}{2\sqrt{\alpha}} \big(\hat{c}_1^\dagger +\bar{z}_1\big) 
- 
\frac{1-\alpha}{2\sqrt{\alpha}} \big(\hat{c}_2 + z_2\big)
\label{eq:20240503-5}, 
\end{equation}
\begin{equation}
\hat{C}^\dagger_{2; 2} 
= 
-\frac{1-\alpha}{2\sqrt{\alpha}} \big(\hat{c}_1 +z_1\big) 
+ 
\frac{1+\alpha}{2\sqrt{\alpha}} \big(\hat{c}_2^\dagger + \bar{z}_2\big),
\label{eq:20240503-6} 
\end{equation}
\begin{equation}
\hat{C}_{2; 1} 
= 
\frac{1+\alpha}{2\sqrt{\alpha}} \big(\hat{c}_1 +z_1\big) 
- 
\frac{1-\alpha}{2\sqrt{\alpha}} \big(\hat{c}_2^\dagger + \bar{z}_2\big)
\label{eq:20240503-7}, 
\end{equation}
\begin{equation}
\hat{C}_{2; 2} 
= 
-\frac{1-\alpha}{2\sqrt{\alpha}} \big(\hat{c}_1^\dagger +\bar{z}_1\big) 
+ 
\frac{1+\alpha}{2\sqrt{\alpha}} \big(\hat{c}_2 + z_2\big),
\label{eq:20240503-8} 
\end{equation}
while $z_i=a_ix_i+ip_i/(2a_i\hbar)$ for $i =1, 2$.
Simultaneously, let us note that in the limit $\alpha\rightarrow 1_{-}$, the above-defined creation operators $\hat{C}_{2; i}^{\dag}$ and the annihilation operators $\hat{C}_{2; i}$ reduce to the form: $\hat{C}^\dagger_{2; 1}=\hat{c}_1^\dagger +\bar{z}_1$, $\hat{C}^\dagger_{2; 2}=\hat{c}_2^\dagger + \bar{z}_2$ oraz $\hat{C}_{2; 1}=\hat{c}_1 +z_1$, $\hat{C}_{2; 2}=\hat{c}_2 + z_2$. 
Based on the determined forms of the creation operators $\hat{C}_{2; i}^{\dag}$ and the annihilation operators $\hat{C}_{2; i}$, for $i=1,2$ expressed by the formulae (\ref{eq:20240503-5}) -- (\ref{eq:20240503-8}) we can transform the Hamiltonian $\hat{H}_2(1,2; z_1,z_2, \xi)$ expressed by the formula (\ref{eq:20240322-2}) to the form
\begin{eqnarray}
\hat{H}_2(1,2; z_1,z_2, \xi) 
&=&
\frac{\hbar}{4\alpha}
\left[(1+\alpha)^2\omega_1+(1-\alpha)^2\omega_2\right]\hat{c}_1^{\dagger}\hat{c}_1
+
\frac{\hbar}{4\alpha}
\left[(1-\alpha)^2\omega_1+(1+\alpha)^2\omega_2\right]
\hat{c}_2^{\dagger}\hat{c}_2
\nonumber\\
&-&
\frac{\hbar}{2\alpha}
(1-\alpha^2)(\omega_1+\omega_2)
\Re\{\left(\hat{c}_1+z_1\right)\left(\hat{c}_2+z_2\right)\}
\nonumber\\
&+&
\frac{\hbar}{2\alpha}
\left[(1-\alpha)^2\omega_1+(1+\alpha)^2\omega_2\right]
\Re\{\bar{z}_2\hat{c}_2\}
\nonumber\\
&+&
\frac{\hbar}{2\alpha}
\left[(1+\alpha)^2\omega_1+(1-\alpha)^2\omega_2\right]
\Re\{\bar{z}_1\hat{c}_1\}
\nonumber\\
&+&
\frac{\hbar}{4\alpha}
\left[(1+\alpha)^2\omega_1+(1-\alpha)^2\omega_2\right]
|z_1|^2
\nonumber\\
&+&
\frac{\hbar}{4\alpha}
\left[(1-\alpha)^2\omega_1+(1+\alpha)^2\omega_2\right)
|z_2|^2
\nonumber\\
&+&
\frac{\hbar}{4\alpha}
(1+\alpha^2)
(\omega_1+\omega_2),
\label{eq:20240503-9}
\end{eqnarray}
where $2\Re\{\bar{z}_i\hat{c}_i\}=\bar{z}_i\hat{c}_i+z_i\hat{c}_i^{\dagger}$. 
Let us note that in the limit $\alpha\rightarrow 1_{-}$, the Hamiltonian~(\ref{eq:20240503-9}) takes the form 
\begin{eqnarray}
\hat{H}_2(1,2; z_1,z_2, 0) 
&=&
\hbar\omega_1\hat{c}_1^{\dagger}\hat{c}_1
+
\hbar\omega_2\hat{c}_2^{\dagger}\hat{c}_2
+
2\hbar\omega_2
\Re\{\bar{z}_2\hat{c}_2\}
+
2\hbar\omega_1
\Re\{\bar{z}_1\hat{c}_1\}
\nonumber\\
&+&
\hbar\omega_1|z_1|^2
+
\hbar\omega_2|z_2|^2
+
\frac{\hbar}{2}(\omega_1+\omega_2).
\label{eq:20240516-02}
\end{eqnarray}
The Hamiltonian given by ~(\ref{eq:20240503-9}) has a rather complex form and, therefore, seems more conveniently expressed using position and momentum operators. 
In order to achieve this goal, we make use of the relationship between the ladder operators $\hat{c}_i^{\dagger}$ and $\hat{c}_i$ associated with the simple harmonic oscillator, and the position $\hat{x}_i$ and momentum operators $\hat{p}_i$, namely 
\begin{equation}
\hat{c}_i^{\dagger}
=
\sqrt{\frac{M\omega_i}{2\hbar}}\hat{x}_i-i\sqrt{\frac{1}{2M\omega_i\hbar}}\hat{p}_i
=
a_i\hat{x}_i-i\frac{1}{2a_i\hbar}\hat{p}_i,
\label{eq:20240503-10}    
\end{equation}

\begin{equation}
\hat{c}_i
=
\sqrt{\frac{M\omega_i}{2\hbar}}\hat{x}_i+i\sqrt{\frac{1}{2M\omega_i\hbar}}\hat{p}_i
=
a_i\hat{x}_i+i\frac{1}{2a_i\hbar}\hat{p}_i.
\label{eq:20240503-11}    
\end{equation}
Hence we can conclude that $\omega_i=(2\hbar/M)a_i^2$, where $a_1=a$ and $a_2=b$ are the previously defined the inverse oscillator lengths along $x_1$ and $x_2$ directions, respectively. 
Accordingly, after expressing the creation operators $\hat{c}_i^{\dagger}$ and the annihilation operators $\hat{c}_i$ in terms of the momentum and the position operators, according to the formulae (\ref{eq:20240503-10}) and (\ref{eq:20240503-11}) in the Hamiltonian (\ref{eq:20240503-9}), and then performing simple but tedious algebraic calculations, we derive the Hamiltonian of our proposed system in the following form 
\begin{eqnarray}
\hat{H}_2(1,2; z_1,z_2, \xi) 
&=&
\frac{(1+\alpha)^2\omega_1+(1-\alpha)^2\omega_2}{4\alpha\omega_1}
\left(
\frac{\hat{p}_1^2}{2M}+\frac{1}{2}M\omega_1^2\hat{x}_1^2 -\frac{1}{2}\hbar\omega_1
\right)
\nonumber\\
&+&
\frac{(1-\alpha)^2\omega_1+(1+\alpha)^2\omega_2}{4\alpha\omega_2}
\left(
\frac{\hat{p}_2^2}{2M}+\frac{1}{2}M\omega_2^2\hat{x}_2^2 -\frac{1}{2}\hbar\omega_2
\right)
\nonumber\\
&+&
\frac{(1-\alpha^2)(\omega_1+\omega_2)}{2\alpha\sqrt{\omega_1\omega_2}}
\left(
\frac{\hat{p}_1\hat{p}_2}{2M}
-
\frac{1}{2}M\omega_1\omega_2\hat{x}_1\hat{x}_2
+
z_{2,i}\sqrt{\frac{\hbar\omega_2}{2M}}\hat{p}_1
+
z_{1,i}\sqrt{\frac{\hbar\omega_1}{2M}}\hat{p}_2
\right)
\nonumber\\
&+&
\frac{(1-\alpha)^2\omega_1+(1+\alpha)^2\omega_2}{2\alpha}
\left(
z_{2,r}\sqrt{\frac{\hbar\omega_2M}{2}}\hat{x}_2
+
z_{2,i}\sqrt{\frac{\hbar}{2\omega_2M}}\hat{p}_2
\right)
\nonumber\\
&+&
\frac{(1+\alpha)^2\omega_1+(1-\alpha)^2\omega_2}{2\alpha}
\left(
z_{1,r}\sqrt{\frac{\hbar\omega_1M}{2}}\hat{x}_1
+
z_{1,i}\sqrt{\frac{\hbar}{2\omega_1M}}\hat{p}_1
\right)
\nonumber\\
&+&
\frac{\hbar}{4\alpha}
\left[(1+\alpha)^2\omega_1+(1-\alpha)^2\omega_2\right]
|z_1|^2
\nonumber\\
&+&
\frac{\hbar}{4\alpha}
\left[(1-\alpha)^2\omega_1+(1+\alpha)^2\omega_2\right)
|z_2|^2
+
\frac{\hbar}{4\alpha}
(1+\alpha^2)
(\omega_1+\omega_2).
\label{eq:20240516-4}
\end{eqnarray}
In the limit $\alpha\rightarrow 1_{-}$, the Hamiltonian~(\ref{eq:20240516-4}) takes the form
\begin{eqnarray}
\hat{H}_2(1,2; z_1,z_2, 0) 
&=&
\frac{\hat{p}_1^2}{2M}+\frac{1}{2}M\omega_1^2\hat{x}_1^2 
+
\hbar\omega_1
\left(
z_{1,r}\sqrt{\frac{M\omega_1}{2\hbar}}\hat{x}_1
+
z_{1,i}\sqrt{\frac{1}{2\hbar\omega_1M}}\hat{p}_1
\right)
+
\hbar\omega_1|z_1|^2
\nonumber\\
&+&
\frac{\hat{p}_2^2}{2M}+\frac{1}{2}M\omega_2^2\hat{x}_2^2 
+
\hbar\omega_2
\left(
z_{2,r}\sqrt{\frac{M\omega_2}{2\hbar}}\hat{x}_2
+
z_{2,i}\sqrt{\frac{1}{2\hbar\omega_2M}}\hat{p}_2
\right)
+
\hbar\omega_2|z_2|^2.
\nonumber\\
\label{eq:20240516-6}    
\end{eqnarray}
Finally, it is still important to note that for the different parameters $a$ and $b$, i.e. $a\neq b$, the wave function, $\widetilde{\psi}_{2,\rm{coh}}^{(\alpha)}(x_1,x_2; z_1,z_2)$, given by (\ref{eq:20231103w06}) lacks a defined symmetry. 
However, if we assume that $a=b$ and simultaneously replace $z_1$ with $z_2$, then the considered wave function is symmetric with respect to the transposition of variables $x_1$ and $x_2$, i.e. the equality $\widetilde{\psi}_{2,\rm{coh}}^{(\alpha)}(x_1,x_2; z_1,z_2)=\widetilde{\psi}_{2,\rm{coh}}^{(\alpha)}(x_2,x_1; z_2,z_1)$ holds. 
Especially, if we assume that $z_1=z_2=0$, the two-particle wave function (\ref{eq:20231103w06}) takes the following form 
\begin{eqnarray}
\widetilde{\psi}_{2,\rm{coh}}^{(\alpha)}(x_1,x_2; 0,0)
&=&
\sqrt{\frac{a^2}{\pi}}
\exp{\left[
-\frac{1+\alpha^2}{4\alpha}a^2x_1^2
-
\frac{1+\alpha^2}{4\alpha}a^2x_2^2
-\frac{1-\alpha^2}{2\alpha}a^2x_1x_2\right]}.
\label{eq:20240517w01}
\end{eqnarray}
As a consequence of the aforementioned assumptions, namely that $a=b$ and $z_1=z_2$, the Hamiltonian~(\ref{eq:20240516-4}) is given the following form
\begin{eqnarray}
\hat{H}_2(1,2; 0,0, \xi) 
&=&
\frac{1+\alpha^2}{2\alpha}
\left(
\frac{\hat{p}_1^2}{2M}+\frac{1}{2}M\Omega^2\hat{x}_1^2 -\frac{1}{2}\hbar\Omega
\right)
+
\frac{1+\alpha^2}{2\alpha}
\left(
\frac{\hat{p}_2^2}{2M}+\frac{1}{2}M\Omega^2\hat{x}_2^2 -\frac{1}{2}\hbar\Omega
\right)
\nonumber\\
&+&
\frac{1-\alpha^2}{\alpha}
\left(
\frac{\hat{p}_1\hat{p}_2}{2M}
-
\frac{1}{2}M\Omega^2\hat{x}_1\hat{x}_2
\right)
+
\frac{1+\alpha^2}{2\alpha}
\hbar\Omega,
\label{eq:20240517-2}
\end{eqnarray}
where $\Omega=\omega_1=\omega_2$ is the frequency of the squeezed isotropic harmonic oscillator. 
Finally, let us note that the form of the obtained Hamiltonian~(\ref{eq:20240517-2}), because of the term proportional to $\hat{p}_1\hat{p}_2$, goes beyond the customarily applied coupled oscillators Hamiltonians, which are used to investigate the entanglement in the system of coupled oscillators~\cite{Gonzalez_SR5p1315y2015,Makarov_PRE97p042203y2018}.
%%%%%%%%%%%%%%%%%%%%%%%%%%%%%%%%%%%%%%%%%%%%%%%%%%%%%%%%%%%%%%%%%%%%%%%%%%%%%%%%%%%%%%%%
\section{\label{sec:5}Conclusions}
%%%%%%%%%%%%%%%%%%%%%%%%%%%%%%%%%%%%%%%%%%%%%%%%%%%%%%%%%%%%%%%%%%%%%%%%%%%%%%%%%%%%%%%%
Following the current consensus in quantum optics, the bipartite squeezed coherent states are generated by applying the squeezing operator acting on the two-mode vacuum state.
In the Bargmann space, this construction may be realized in at least two ways, depending on the choice of linear combination coefficients with respect to the established basis.
In this study, we consider two sets of linear combination coefficients: one comprising the products of the holomorphic Hermite polynomials of one complex variable, and the other is formed from the holomorphic Hermite polynomials of two complex variables.
These coefficients can be regarded as the result of acting on the unimodal or bipartite realizations of the $\mathfrak{su}(1,1)$ algebra on the basis formed from the products of monomials in the Bargmann space.
By applying the Segal-Bargmann transform, we go to the Schr{\"o}dinger space, wherein the wave functions corresponding to the bipartite squeezed coherent states are expressed by the Gaussian functions. 
These functions take a factorizable or non-factorizable form depending on the method of squeezing employed. 
This naturally gives rise to the question of whether the states thus obtained are entangled. 
To address this question, we have employed the Horodecki-Peres criterion for continuous-variable systems adapted to the phase-space approach.
As demonstrated by Simon, the realization of this criterion in the phase space has a straightforward interpretation in terms of the Wigner distribution functions. 
This is because it has been sufficient to limit our examination to determine the symplectic eigenvalues of the appropriately modified covariance matrix of the Gaussians related to the Wigner distribution functions, following Simon's recipe.
The result we have obtained explicitly depends on the value of the squeezing parameter, $\xi=-(\log{\alpha})/2$, where the parameter $\alpha\in(0,1)$.
Furthermore, we have demonstrated that applying the quantitative log-negativity criterion for the entanglement allows us to state that the no squeezing $\xi=0$ is equivalent to the absence of the entanglement in the examined states. 
Conversely, when the squeezing parameter $\xi\neq 0$, the states are entangled, wherein the maximal entanglement, exemplified by the EPR states, occurs when the parameter value approaches infinity. 
One of the central results of our findings presented in this paper is the following observation. 
For the families of bipartite Gaussian functions considered in this paper, we conclude, based on the Simon-Horodecki-Peres criterion, that the considered Gaussian wave functions expressed in the originally defined position coordinates, denoted by $x_1$ and $x_2$, are factorizable if and only if the corresponding state is separable. 
It indicates that the non-factorizability of the considered Gaussian wave functions expressed in these coordinates is equivalent to the entanglement of the corresponding state.  
Let us also note that, admittedly, a point transformation of the original position coordinates $(x_1,x_2)$, such as a centre of mass and relative coordinates, allows the examined Gaussian wave function to be factored. 
However, this does not imply the separability of the corresponding state. 
In order to demonstrate the findings of our study, we have proposed and investigated a theoretical model of a two-particle mechanical system that allows expounding physical interpretation of the presented result. 
The Hamiltonian we have derived describe a system consisting of two coupled anisotropic harmonic oscillators depending on the parameter $\alpha$.
We have demonstrated that the limit form that corresponds to $\alpha\rightarrow 1_{-}$ or alternatively $\xi\to 0_{+}$ of the aforementioned Hamiltonian reduces to the Hamiltonian of the two independent anisotropic harmonic oscillators shifted in positions and momenta.
Furthermore, we also have shown that assuming no shifts and assuming identical oscillation frequencies, the Hamiltonian we have obtained is consistent with the simplified model discussed in the previous work. 
Based on the results discussed in this paper, it can also be concluded that performing a unitary transformation corresponding to a canonical transformation on phase-space variables can lead to the decoupling of a system of interacting isotropic harmonic oscillators. 
However, the result obtained in this way leads to the loss of the individual characteristics of the input subsystems.  
Furthermore, each of the subsystems obtained in this way is based on new phase-space variables that mix the original variables used to describe the input systems.

%%%%%%%%%%%%%%%%%%%%%%%%%%%%%%%%%%%%%%%%%%%%%%%%%%%%%%%%%%%%%%%%%%%%%%%%%%%%%%%%%%%%%%%%
\appendix
\section{\label{app:Mehler} Appendix: Mehler formulae}
%%%%%%%%%%%%%%%%%%%%%%%%%%%%%%%%%%%%%%%%%%%%%%%%%%%%%%%%%%%%%%%%%%%%%%%%%%%%%%%%%%%%%%%%
\begin{lemma}
Let $z_1,z_2\in\mathbb{C}$ and $|t|<1$. 
The exponential generating function for the product of two single vaiables holomorphic Hermite polynomials has the form
\begin{equation}
%\fl
\sum_{l=0}^{\infty}
\frac{t^l}{2^ll!}H_l(z_1)H_l(z_2)
=
\frac{1}{\sqrt{1-t^2}}
\exp{\left[
\frac{2tz_1z_2-t^2(z_1^2+z_2^2)}{1-t^2}
\right]}.
\label{eq:20231103w03}
\end{equation}    
\end{lemma}
\noindent
{\em Proof.} 
To prove the Mehler formula~(\ref{eq:20231103w03}) we apply the operational calculus~\cite{RomanS}. 
According to this we introduce the operator form of the Hermite polynomials, namely $\exp[-(1/4)\partial_x^2] (2 x)^l = H_l(x)$. 
The left-hand side (LHS) of Eq.~(\ref{eq:20231103w03}) reads
\begin{equation}
\sum_{l=0}^{\infty}
\frac{t^l}{2^ll!}H_l(z_1)H_l(z_2)
    = \exp\left(-\frac{1}{4}\partial_{z_1}^2\right) \exp\left(-\frac{1}{4}\partial_{z_2}^2\right) \exp(2tz_1 z_2).
    \label{eq:20240719w01}
\end{equation}
The action of the operator $\exp\left[-(1/4)\partial_{z_2}^2\right]$ on the exponential function $\exp(2tz_1 z_2)$ is equal to $\exp(-t^2 z_1^2 + 2 t z_1 z_2)=\exp[-(t z_1 - z_2)^2 + z_2^2]$. 
Then, setting $u = t z_1 - z_2$, allows one to present $\exp[-(1/4) \partial_{z_1}^2]\exp[-(t z_1 - z_2)^2 + z_2^2]$ in the form
\begin{equation}\label{eq:20240718w01}
    \text{\rm LHS\,\, of\,\, Eq. (\ref{eq:20231103w03})} 
        =  \exp(z_2^2) \exp\left(-\frac{t^2}{4} \partial_u^2\right) \exp(-u^2) =  \frac{\exp(z_2^2)}{\sqrt{1-t^2}} \exp\left(-\frac{u^2}{1-t^2}\right),
\end{equation}
which, after coming back to variables $z_1$ and $z_2$, gives Eq.~(\ref{eq:20231103w03}).

\begin{lemma}
Let $z_1,z_2\in\mathbb{C}$, $u,v\in\mathbb{R}$, and $|st|<1$. 
The exponential generating function for the product of holomorphic Hermite polynomials of two complex variables and two Hermite polynomials of one real variable has the form
\begin{eqnarray}
&&
%\fl
\sum_{m=0}^\infty 
\sum_{n=0}^\infty 
\frac{s^m t^n}{\sqrt{2^{m+n}}\,m!n!}
{\rm H}_{m,n}(z_1,z_2){\rm H}_{m}(u) {\rm H}_n(v) 
= 
\frac{1}{\sqrt{1-s^2t^2}} 
\nonumber\\ 
&\times&
\exp\left[\frac{2\sqrt{2}\left(suz_1 + tvz_2
+st\left(svz_1 + tuz_2\right)\right)-s^2z_1^2-t^2z_2^2}{2\left(1-s^2t^2\right)}\right]
\nonumber\\
&\times&
\exp\left[\frac{-4stuv-2s^2t^2\left(z_1 z_2
+u^2+v^2\right)}{2\left(1-s^2t^2\right)}\right] .
\label{eq:20231103w05}
\end{eqnarray}
\end{lemma}

\noindent
{\em Proof.} 
To prove Mehler formula~(\ref{eq:20231103w05}), we recall the generating function of two complex variables Hermite polynomials
\begin{equation}
\sum_{m=0}^\infty \sum_{n=0}^\infty \frac{s^m t^n}{m! n!} H_{m, n}(z_1, _2) 
= 
\exp(s z_1 + t z_2 - st),
\label{eq:20240719w01-a}  
\end{equation}
and use once more the operator form of the Hermite polynomials, we get
\begin{eqnarray}
\sum_{m=0}^\infty 
\sum_{n=0}^\infty 
&\;&
\frac{s^m t^n}{\sqrt{2^{m+n}}\,m!n!}
{\rm H}_{m,n}(z_1,z_2){\rm H}_{m}(u) {\rm H}_n(v) 
\nonumber\\
&=&
e^{-\frac{1}{4}(\partial_u^2 + \partial_v^2)} 
\sum_{m=0}^\infty
\sum_{n=0}^\infty 
\frac{(\sqrt{2} u s)^m (\sqrt{2} v t)^n}{\,m!n!}
{\rm H}_{m,n}(z_1,z_2)
\nonumber\\
&=&
e^{-\frac{1}{4}(\partial_u^2 + \partial_v^2)} 
\exp{\left[\sqrt{2}us z_1+\sqrt{2}vt z_2 - 2usvt\right]}.
\label{eq:20240719-001}
\end{eqnarray}
Explicit calculation of the derivatives in expression~(\ref{eq:20240719-001})  completes the proof. 

%\bibliography{quant-bib.bib} 

\begin{thebibliography}{61}%
\makeatletter
\providecommand \@ifxundefined [1]{%
 \@ifx{#1\undefined}
}%
\providecommand \@ifnum [1]{%
 \ifnum #1\expandafter \@firstoftwo
 \else \expandafter \@secondoftwo
 \fi
}%
\providecommand \@ifx [1]{%
 \ifx #1\expandafter \@firstoftwo
 \else \expandafter \@secondoftwo
 \fi
}%
\providecommand \natexlab [1]{#1}%
\providecommand \enquote  [1]{``#1''}%
\providecommand \bibnamefont  [1]{#1}%
\providecommand \bibfnamefont [1]{#1}%
\providecommand \citenamefont [1]{#1}%
\providecommand \href@noop [0]{\@secondoftwo}%
\providecommand \href [0]{\begingroup \@sanitize@url \@href}%
\providecommand \@href[1]{\@@startlink{#1}\@@href}%
\providecommand \@@href[1]{\endgroup#1\@@endlink}%
\providecommand \@sanitize@url [0]{\catcode `\\12\catcode `\$12\catcode
  `\&12\catcode `\#12\catcode `\^12\catcode `\_12\catcode `\%12\relax}%
\providecommand \@@startlink[1]{}%
\providecommand \@@endlink[0]{}%
\providecommand \url  [0]{\begingroup\@sanitize@url \@url }%
\providecommand \@url [1]{\endgroup\@href {#1}{\urlprefix }}%
\providecommand \urlprefix  [0]{URL }%
\providecommand \Eprint [0]{\href }%
\providecommand \doibase [0]{http://dx.doi.org/}%
\providecommand \selectlanguage [0]{\@gobble}%
\providecommand \bibinfo  [0]{\@secondoftwo}%
\providecommand \bibfield  [0]{\@secondoftwo}%
\providecommand \translation [1]{[#1]}%
\providecommand \BibitemOpen [0]{}%
\providecommand \bibitemStop [0]{}%
\providecommand \bibitemNoStop [0]{.\EOS\space}%
\providecommand \EOS [0]{\spacefactor3000\relax}%
\providecommand \BibitemShut  [1]{\csname bibitem#1\endcsname}%
\let\auto@bib@innerbib\@empty
%</preamble>
\bibitem [{\citenamefont {Zhang}\ \emph {et~al.}(1990)\citenamefont {Zhang},
  \citenamefont {Feng},\ and\ \citenamefont {Gilmore}}]{Zhang_RMP62p867y1990}%
  \BibitemOpen
  \bibfield  {author} {\bibinfo {author} {\bibfnamefont {W.-M.}\ \bibnamefont
  {Zhang}}, \bibinfo {author} {\bibfnamefont {D.~H.}\ \bibnamefont {Feng}}, \
  and\ \bibinfo {author} {\bibfnamefont {R.}~\bibnamefont {Gilmore}},\ }\href
  {\doibase 10.1103/RevModPhys.62.867} {\bibfield  {journal} {\bibinfo
  {journal} {Rev. Mod. Phys.}\ }\textbf {\bibinfo {volume} {62}},\ \bibinfo
  {pages} {867} (\bibinfo {year} {1990})}\BibitemShut {NoStop}%
\bibitem [{\citenamefont {Gazeau}(2009)}]{JPGazeau09}%
  \BibitemOpen
  \bibfield  {author} {\bibinfo {author} {\bibfnamefont {J.-P.}\ \bibnamefont
  {Gazeau}},\ }\href {\doibase 10.1002/9783527628285} {\emph {\bibinfo {title}
  {Coherent States in Quantum Physics}}}\ (\bibinfo  {publisher} {Wiley-VCH},\
  \bibinfo {year} {2009})\BibitemShut {NoStop}%
\bibitem [{\citenamefont {Klauder}\ and\ \citenamefont
  {Skagerstam}(1985)}]{JRKlauder85}%
  \BibitemOpen
  \bibfield  {author} {\bibinfo {author} {\bibfnamefont {J.~R.}\ \bibnamefont
  {Klauder}}\ and\ \bibinfo {author} {\bibfnamefont {B.~S.}\ \bibnamefont
  {Skagerstam}},\ }\href {\doibase 10.1142/0096} {\emph {\bibinfo {title}
  {Coherent states. Applications in Physics and Mathematical Physics}}}\
  (\bibinfo  {publisher} {World Scientific},\ \bibinfo {year}
  {1985})\BibitemShut {NoStop}%
\bibitem [{\citenamefont {Schleich}(2001)}]{Schleich_book2001}%
  \BibitemOpen
  \bibfield  {author} {\bibinfo {author} {\bibfnamefont {W.~P.}\ \bibnamefont
  {Schleich}},\ }\href {\doibase 10.1002/3527602976} {\emph {\bibinfo {title}
  {{Quantum Optics in Phase Space}}}}\ (\bibinfo  {publisher} {John Wiley and
  Sons, Inc.},\ \bibinfo {address} {New York},\ \bibinfo {year}
  {2001})\BibitemShut {NoStop}%
\bibitem [{\citenamefont {Gerry}\ and\ \citenamefont
  {Knight}(2004)}]{Gerry_book2004}%
  \BibitemOpen
  \bibfield  {author} {\bibinfo {author} {\bibfnamefont {C.}~\bibnamefont
  {Gerry}}\ and\ \bibinfo {author} {\bibfnamefont {P.}~\bibnamefont {Knight}},\
  }\href {\doibase 10.1017/CBO9780511791239} {\emph {\bibinfo {title}
  {{Introductory Quantum Optics}}}}\ (\bibinfo  {publisher} {Cambridge
  University Press},\ \bibinfo {year} {2004})\BibitemShut {NoStop}%
\bibitem [{\citenamefont {Arfken}\ \emph {et~al.}(2013)\citenamefont {Arfken},
  \citenamefont {Weber},\ and\ \citenamefont {Harris}}]{Arfken}%
  \BibitemOpen
  \bibfield  {author} {\bibinfo {author} {\bibfnamefont {G.~B.}\ \bibnamefont
  {Arfken}}, \bibinfo {author} {\bibfnamefont {H.~J.}\ \bibnamefont {Weber}}, \
  and\ \bibinfo {author} {\bibfnamefont {F.~E.}\ \bibnamefont {Harris}},\
  }\href {\doibase 10.1016/C2009-0-30629-7} {\emph {\bibinfo {title}
  {Mathematical Methods for Physicists. {A} Comprehensive Guide}}},\ \bibinfo
  {edition} {seventh}\ ed.\ (\bibinfo  {publisher} {Academic Press},\ \bibinfo
  {year} {2013})\BibitemShut {NoStop}%
\bibitem [{\citenamefont {Scully}\ and\ \citenamefont
  {Zubairy}(1997)}]{MOSculy1997}%
  \BibitemOpen
  \bibfield  {author} {\bibinfo {author} {\bibfnamefont {A.~O.}\ \bibnamefont
  {Scully}}\ and\ \bibinfo {author} {\bibfnamefont {S.}~\bibnamefont
  {Zubairy}},\ }\href {\doibase 10.1017/CBO9780511813993} {\emph {\bibinfo
  {title} {Quantum Optics}}}\ (\bibinfo  {publisher} {Cambridge University
  Press},\ \bibinfo {year} {1997})\BibitemShut {NoStop}%
\bibitem [{\citenamefont {Klauder}(1960)}]{Klauder1}%
  \BibitemOpen
  \bibfield  {author} {\bibinfo {author} {\bibfnamefont {J.~R.}\ \bibnamefont
  {Klauder}},\ }\href {\doibase https://doi.org/10.1016/0003-4916(60)90131-7}
  {\bibfield  {journal} {\bibinfo  {journal} {Ann. Phys.}\ }\textbf {\bibinfo
  {volume} {11}},\ \bibinfo {pages} {123} (\bibinfo {year} {1960})}\BibitemShut
  {NoStop}%
\bibitem [{\citenamefont {Klauder}(1963{\natexlab{a}})}]{Klauder2}%
  \BibitemOpen
  \bibfield  {author} {\bibinfo {author} {\bibfnamefont {J.~R.}\ \bibnamefont
  {Klauder}},\ }\href {\doibase https://doi.org/10.1063/1.1704035} {\bibfield
  {journal} {\bibinfo  {journal} {J. Math. Phys.}\ }\textbf {\bibinfo {volume}
  {4}},\ \bibinfo {pages} {1058–1073} (\bibinfo {year}
  {1963}{\natexlab{a}})}\BibitemShut {NoStop}%
\bibitem [{\citenamefont {Klauder}(1963{\natexlab{b}})}]{JKlauder63}%
  \BibitemOpen
  \bibfield  {author} {\bibinfo {author} {\bibfnamefont {J.}~\bibnamefont
  {Klauder}},\ }\href {\doibase https://doi.org/10.1063/1.1704034} {\bibfield
  {journal} {\bibinfo  {journal} {J. Math. Phys.}\ }\textbf {\bibinfo {volume}
  {4}},\ \bibinfo {pages} {1055} (\bibinfo {year}
  {1963}{\natexlab{b}})}\BibitemShut {NoStop}%
\bibitem [{\citenamefont {Parthasarathy}(1992)}]{KRParthasarathy92}%
  \BibitemOpen
  \bibfield  {author} {\bibinfo {author} {\bibfnamefont {K.~R.}\ \bibnamefont
  {Parthasarathy}},\ }\href {\doibase 10.1007/978-3-0348-8641-3} {\emph
  {\bibinfo {title} {{An introduction to quantum stochastic calculus}}}}\
  (\bibinfo  {publisher} {Springer Basel AG},\ \bibinfo {year}
  {1992})\BibitemShut {NoStop}%
\bibitem [{\citenamefont {Gazeau}(2019)}]{JPGazeau19}%
  \BibitemOpen
  \bibfield  {author} {\bibinfo {author} {\bibfnamefont {J.-P.}\ \bibnamefont
  {Gazeau}},\ }in\ \href {\doibase 10.1007/978-3-030-20087-9} {\emph {\bibinfo
  {booktitle} {Integrability, Supersymmetry and Coherent States. A Volume in
  Honour of {P}rofessor {V}{\'e}ronique {H}ussin}}}\ (\bibinfo  {publisher}
  {Springer Cham},\ \bibinfo {year} {2019})\ pp.\ \bibinfo {pages}
  {69--101}\BibitemShut {NoStop}%
\bibitem [{\citenamefont {Dodonov}(2002)}]{Dodonov_JOBQSO4pR1y2002}%
  \BibitemOpen
  \bibfield  {author} {\bibinfo {author} {\bibfnamefont {V.~V.}\ \bibnamefont
  {Dodonov}},\ }\href {\doibase 10.1088/1464-4266/4/1/201} {\bibfield
  {journal} {\bibinfo  {journal} {J. Opt. B: Quantum Semiclass Opt.}\ }\textbf
  {\bibinfo {volume} {4}},\ \bibinfo {pages} {R1} (\bibinfo {year}
  {2002})}\BibitemShut {NoStop}%
\bibitem [{\citenamefont {Gazeau}\ \emph {et~al.}(2022)\citenamefont {Gazeau},
  \citenamefont {Hussin}, \citenamefont {Moran},\ and\ \citenamefont
  {Zelaya}}]{JPGazeau22}%
  \BibitemOpen
  \bibfield  {author} {\bibinfo {author} {\bibfnamefont {J.-P.}\ \bibnamefont
  {Gazeau}}, \bibinfo {author} {\bibfnamefont {V.}~\bibnamefont {Hussin}},
  \bibinfo {author} {\bibfnamefont {J.}~\bibnamefont {Moran}}, \ and\ \bibinfo
  {author} {\bibfnamefont {K.}~\bibnamefont {Zelaya}},\ }\href {\doibase
  10.1016/j.aop.2022.168888} {\bibfield  {journal} {\bibinfo  {journal} {Ann.
  Phys. (N.Y)}\ }\textbf {\bibinfo {volume} {441}},\ \bibinfo {pages} {168888}
  (\bibinfo {year} {2022})}\BibitemShut {NoStop}%
\bibitem [{\citenamefont {Bergeron}\ \emph
  {et~al.}(2024{\natexlab{a}})\citenamefont {Bergeron}, \citenamefont {Gazeau},
  \citenamefont {Ma{\l}kiewicz},\ and\ \citenamefont {Peter}}]{HBergeron24}%
  \BibitemOpen
  \bibfield  {author} {\bibinfo {author} {\bibfnamefont {H.}~\bibnamefont
  {Bergeron}}, \bibinfo {author} {\bibfnamefont {J.-P.}\ \bibnamefont
  {Gazeau}}, \bibinfo {author} {\bibfnamefont {P.}~\bibnamefont
  {Ma{\l}kiewicz}}, \ and\ \bibinfo {author} {\bibfnamefont {P.}~\bibnamefont
  {Peter}},\ }\href {\doibase 10.1103/PhysRevD.109.023516} {\bibfield
  {journal} {\bibinfo  {journal} {Phys. Rev. D}\ }\textbf {\bibinfo {volume}
  {109}},\ \bibinfo {pages} {023516} (\bibinfo {year}
  {2024}{\natexlab{a}})}\BibitemShut {NoStop}%
\bibitem [{\citenamefont {Bergeron}\ \emph
  {et~al.}(2024{\natexlab{b}})\citenamefont {Bergeron}, \citenamefont
  {Ma{\l}kiewicz},\ and\ \citenamefont {Peter}}]{HBergeron24a}%
  \BibitemOpen
  \bibfield  {author} {\bibinfo {author} {\bibfnamefont {H.}~\bibnamefont
  {Bergeron}}, \bibinfo {author} {\bibfnamefont {P.}~\bibnamefont
  {Ma{\l}kiewicz}}, \ and\ \bibinfo {author} {\bibfnamefont {P.}~\bibnamefont
  {Peter}},\ }\href {\doibase 10.1103/PhysRevD.110.043512} {\bibfield
  {journal} {\bibinfo  {journal} {Phys. Rev. D}\ }\textbf {\bibinfo {volume}
  {110}},\ \bibinfo {pages} {043512} (\bibinfo {year}
  {2024}{\natexlab{b}})}\BibitemShut {NoStop}%
\bibitem [{\citenamefont {Gazeau}\ and\ \citenamefont
  {Pejhan}(2023)}]{JPGazeau23}%
  \BibitemOpen
  \bibfield  {author} {\bibinfo {author} {\bibfnamefont {J.-P.}\ \bibnamefont
  {Gazeau}}\ and\ \bibinfo {author} {\bibfnamefont {H.}~\bibnamefont
  {Pejhan}},\ }\href {\doibase 10.1103/PhysRevD.108.065012} {\bibfield
  {journal} {\bibinfo  {journal} {Phys. Rev. D}\ }\textbf {\bibinfo {volume}
  {108}},\ \bibinfo {pages} {065012} (\bibinfo {year} {2023})}\BibitemShut
  {NoStop}%
\bibitem [{\citenamefont {G\'{o}rska}\ and\ \citenamefont
  {Horzela}(2023)}]{KGorska23}%
  \BibitemOpen
  \bibfield  {author} {\bibinfo {author} {\bibfnamefont {K.}~\bibnamefont
  {G\'{o}rska}}\ and\ \bibinfo {author} {\bibfnamefont {A.}~\bibnamefont
  {Horzela}},\ }\href {\doibase 10.1088/1742-6596/2667/1/012008} {\bibfield
  {journal} {\bibinfo  {journal} {J. Phys. Conf. Ser.A}\ }\textbf {\bibinfo
  {volume} {2667}},\ \bibinfo {pages} {012008} (\bibinfo {year}
  {2023})}\BibitemShut {NoStop}%
\bibitem [{\citenamefont {Thirring}\ \emph {et~al.}(2011)\citenamefont
  {Thirring}, \citenamefont {Bertlmanna}, \citenamefont {K\"{o}hler},\ and\
  \citenamefont {Narnhofer}}]{Thirring_EPJD64p181y2011}%
  \BibitemOpen
  \bibfield  {author} {\bibinfo {author} {\bibfnamefont {W.}~\bibnamefont
  {Thirring}}, \bibinfo {author} {\bibfnamefont {R.~A.}\ \bibnamefont
  {Bertlmanna}}, \bibinfo {author} {\bibfnamefont {P.}~\bibnamefont
  {K\"{o}hler}}, \ and\ \bibinfo {author} {\bibfnamefont {H.}~\bibnamefont
  {Narnhofer}},\ }\href {\doibase 10.1140/epjd/e2011-20452-1} {\bibfield
  {journal} {\bibinfo  {journal} {Eur. Phys. J. D}\ }\textbf {\bibinfo {volume}
  {64}},\ \bibinfo {pages} {181} (\bibinfo {year} {2011})}\BibitemShut
  {NoStop}%
\bibitem [{\citenamefont {de~la Torre}\ \emph {et~al.}(2010)\citenamefont
  {de~la Torre}, \citenamefont {Goyeneche},\ and\ \citenamefont
  {Leitao}}]{Torre_EJP31p325y2010}%
  \BibitemOpen
  \bibfield  {author} {\bibinfo {author} {\bibfnamefont {A.~C.}\ \bibnamefont
  {de~la Torre}}, \bibinfo {author} {\bibfnamefont {D.}~\bibnamefont
  {Goyeneche}}, \ and\ \bibinfo {author} {\bibfnamefont {L.}~\bibnamefont
  {Leitao}},\ }\href {\doibase 10.1088/0143-0807/31/2/010} {\bibfield
  {journal} {\bibinfo  {journal} {Eur. J. Phys.}\ }\textbf {\bibinfo {volume}
  {31}},\ \bibinfo {pages} {325} (\bibinfo {year} {2010})}\BibitemShut
  {NoStop}%
\bibitem [{\citenamefont {G\'{o}rska}\ \emph {et~al.}(2018)\citenamefont
  {G\'{o}rska}, \citenamefont {Horzela},\ and\ \citenamefont
  {Szafraniec}}]{lumini}%
  \BibitemOpen
  \bibfield  {author} {\bibinfo {author} {\bibfnamefont {K.}~\bibnamefont
  {G\'{o}rska}}, \bibinfo {author} {\bibfnamefont {A.}~\bibnamefont {Horzela}},
  \ and\ \bibinfo {author} {\bibfnamefont {F.~H.}\ \bibnamefont {Szafraniec}},\
  }in\ \href {\doibase https://doi.org/10.1007/978-3-319-76732-1\_5} {\emph
  {\bibinfo {booktitle} {Coherent States and Their Applications}}},\ Vol.\
  \bibinfo {volume} {205}\ (\bibinfo  {publisher} {Springer},\ \bibinfo {year}
  {2018})\ pp.\ \bibinfo {pages} {89--117}\BibitemShut {NoStop}%
\bibitem [{\citenamefont {Schumaker}(1986)}]{Schumaker_PR135p317y1986}%
  \BibitemOpen
  \bibfield  {author} {\bibinfo {author} {\bibfnamefont {B.~L.}\ \bibnamefont
  {Schumaker}},\ }\href {\doibase https://doi.org/10.1016/0370-1573(86)90179-1}
  {\bibfield  {journal} {\bibinfo  {journal} {Phys. Rep.}\ }\textbf {\bibinfo
  {volume} {135}},\ \bibinfo {pages} {317} (\bibinfo {year}
  {1986})}\BibitemShut {NoStop}%
% \bibitem [{\citenamefont {Barnett}\ and\ \citenamefont
%   {L.}(1987)}]{SMBarnett87}%
%   \BibitemOpen
%   \bibfield  {author} {\bibinfo {author} {\bibfnamefont {S.~M.}\ \bibnamefont
%   {Barnett}}\ and\ \bibinfo {author} {\bibfnamefont {K.~P.}\ \bibnamefont
%   {L.}},\ }\href {\doibase 10.1080/09500348714550781} {\bibfield  {journal}
%   {\bibinfo  {journal} {J. Mod. Opt.}\ }\textbf {\bibinfo {volume} {34}},\
%   \bibinfo {pages} {841} (\bibinfo {year} {1987})}\BibitemShut {NoStop}%
\bibitem [{\citenamefont {Barnett}\ and\ \citenamefont
  {Knight}(1987)}]{SMBarnett87}%
  \BibitemOpen
  \bibfield  {author} {\bibinfo {author} {\bibfnamefont {S.~M.}\ \bibnamefont
  {Barnett}}\ and\ \bibinfo {author} {\bibfnamefont {P.~L.}\ \bibnamefont
  {Knight}},\ }\href {\doibase 10.1080/09500348714550781} {\bibfield  {journal}
  {\bibinfo  {journal} {J. Mod. Opt.}\ }\textbf {\bibinfo {volume} {34}},\
  \bibinfo {pages} {841} (\bibinfo {year} {1987})}\BibitemShut {NoStop}%
\bibitem [{\citenamefont {Bishop}\ and\ \citenamefont
  {Vourdas}(1988)}]{RFBishop88}%
  \BibitemOpen
  \bibfield  {author} {\bibinfo {author} {\bibfnamefont {R.~F.}\ \bibnamefont
  {Bishop}}\ and\ \bibinfo {author} {\bibfnamefont {A.}~\bibnamefont
  {Vourdas}},\ }\href {\doibase 10.1007/BF01313941} {\bibfield  {journal}
  {\bibinfo  {journal} {Z. Phys. B - Condensed Matter}\ }\textbf {\bibinfo
  {volume} {71}},\ \bibinfo {pages} {527} (\bibinfo {year} {1988})}\BibitemShut
  {NoStop}%
\bibitem [{\citenamefont {Vourdas}(1992)}]{AVourdas92}%
  \BibitemOpen
  \bibfield  {author} {\bibinfo {author} {\bibfnamefont {A.}~\bibnamefont
  {Vourdas}},\ }\href {\doibase 10.1103/PhysRevA.46.442} {\bibfield  {journal}
  {\bibinfo  {journal} {Phys. Rev. A}\ }\textbf {\bibinfo {volume} {46}},\
  \bibinfo {pages} {442} (\bibinfo {year} {1992})}\BibitemShut {NoStop}%
\bibitem [{\citenamefont {Ekert}\ and\ \citenamefont
  {Knight}(1989)}]{AKEkert89}%
  \BibitemOpen
  \bibfield  {author} {\bibinfo {author} {\bibfnamefont {A.~K.}\ \bibnamefont
  {Ekert}}\ and\ \bibinfo {author} {\bibfnamefont {P.~L.}\ \bibnamefont
  {Knight}},\ }\href {\doibase 10.1119/1.15922} {\bibfield  {journal} {\bibinfo
   {journal} {Am. J. Phys.}\ }\textbf {\bibinfo {volume} {57}},\ \bibinfo
  {pages} {692} (\bibinfo {year} {1989})}\BibitemShut {NoStop}%
\bibitem [{\citenamefont {Hong-Yi}(1990)}]{FHong-yi90}%
  \BibitemOpen
  \bibfield  {author} {\bibinfo {author} {\bibfnamefont {F.}~\bibnamefont
  {Hong-Yi}},\ }\href {\doibase 10.1103/PhysRevA.41.1526} {\bibfield  {journal}
  {\bibinfo  {journal} {Phys. Rev. A}\ }\textbf {\bibinfo {volume} {41}},\
  \bibinfo {pages} {1526} (\bibinfo {year} {1990})}\BibitemShut {NoStop}%
\bibitem [{\citenamefont {Yeoman}\ and\ \citenamefont
  {Barnett}(1993)}]{GYeoman93}%
  \BibitemOpen
  \bibfield  {author} {\bibinfo {author} {\bibfnamefont {G.}~\bibnamefont
  {Yeoman}}\ and\ \bibinfo {author} {\bibfnamefont {S.~M.}\ \bibnamefont
  {Barnett}},\ }\href {\doibase 10.1080/09500349314551561} {\bibfield
  {journal} {\bibinfo  {journal} {J. Mod. Opt.}\ }\textbf {\bibinfo {volume}
  {40}},\ \bibinfo {pages} {1497} (\bibinfo {year} {1993})}\BibitemShut
  {NoStop}%
\bibitem [{\citenamefont {Hong-yi}\ and\ \citenamefont
  {Yue}(1996)}]{FHong-yi96}%
  \BibitemOpen
  \bibfield  {author} {\bibinfo {author} {\bibfnamefont {F.}~\bibnamefont
  {Hong-Yi}}\ and\ \bibinfo {author} {\bibfnamefont {F.}~\bibnamefont {Yue}},\
  }\href {\doibase 10.1103/PhysRevA.54.958} {\bibfield  {journal} {\bibinfo
  {journal} {Phys. Rev. A}\ }\textbf {\bibinfo {volume} {54}},\ \bibinfo
  {pages} {958} (\bibinfo {year} {1996})}\BibitemShut {NoStop}%
\bibitem [{\citenamefont {Bagarello}(2022)}]{FBagarello22}%
  \BibitemOpen
  \bibfield  {author} {\bibinfo {author} {\bibfnamefont {F.}~\bibnamefont
  {Bagarello}},\ }\href {\doibase 10.1007/978-3-030-94999-0} {\emph {\bibinfo
  {title} {Mathematical Methods for Physicists. {A} Comprehensive
  GuidePseudo-Bosons and Their Coherent States}}}\ (\bibinfo  {publisher}
  {Springer},\ \bibinfo {year} {2022})\BibitemShut {NoStop}%
\bibitem [{\citenamefont {Chru\'{s}ci\'{n}ski}(2004)}]{DChruscinski2004}%
  \BibitemOpen
  \bibfield  {author} {\bibinfo {author} {\bibfnamefont {D.}~\bibnamefont
  {Chru\'{s}ci\'{n}ski}},\ }\href {\doibase
  https://doi.org/10.1016/j.physleta.2004.05.046} {\bibfield  {journal}
  {\bibinfo  {journal} {Phys. Lett.}\ }\textbf {\bibinfo {volume} {327}},\
  \bibinfo {pages} {290} (\bibinfo {year} {2004})}\BibitemShut {NoStop}%
\bibitem [{\citenamefont {Hall}(2000)}]{BCHall2000}%
  \BibitemOpen
  \bibfield  {author} {\bibinfo {author} {\bibfnamefont {B.~C.}\ \bibnamefont
  {Hall}},\ }\href {\doibase
  https://citeseerx.ist.psu.edu/document?repid=rep1&type=pdf&doi=16985b428d3a3d496134534e48c9189c4eaed574}
  {\bibfield  {journal} {\bibinfo  {journal} {Contemp. Math.}\ }\textbf
  {\bibinfo {volume} {260}},\ \bibinfo {pages} {1} (\bibinfo {year}
  {2000})}\BibitemShut {NoStop}%
\bibitem [{\citenamefont {Vourdas}(2006)}]{AVourdas2006}%
  \BibitemOpen
  \bibfield  {author} {\bibinfo {author} {\bibfnamefont {A.}~\bibnamefont
  {Vourdas}},\ }\href {\doibase 10.1088/0305-4470/39/7/R01} {\bibfield
  {journal} {\bibinfo  {journal} {J. Phys. A: Math. Gen.}\ }\textbf {\bibinfo
  {volume} {39}},\ \bibinfo {pages} {R65} (\bibinfo {year} {2006})}\BibitemShut
  {NoStop}%
\bibitem [{\citenamefont {Bargmann}(1961)}]{VBargmann61}%
  \BibitemOpen
  \bibfield  {author} {\bibinfo {author} {\bibfnamefont {V.}~\bibnamefont
  {Bargmann}},\ }\href {\doibase https://doi.org/10.1002/cpa.3160140303}
  {\bibfield  {journal} {\bibinfo  {journal} {Commun. Pure Appl. Math.}\
  }\textbf {\bibinfo {volume} {14}},\ \bibinfo {pages} {187} (\bibinfo {year}
  {1961})}\BibitemShut {NoStop}%
% \bibitem [{\citenamefont {van Eijndhoven}\ and\ \citenamefont
%   {H.}(1990)}]{SLLvanEijndhoven90}%
%   \BibitemOpen
%   \bibfield  {author} {\bibinfo {author} {\bibfnamefont {S.~J.~L.}\
%   \bibnamefont {van Eijndhoven}}\ and\ \bibinfo {author} {\bibfnamefont
%   {M.~J.~L.}\ \bibnamefont {H.}},\ }\href {\doibase
%   https://doi.org/10.1016/0022-247X(90)90334-C} {\bibfield  {journal} {\bibinfo
%    {journal} {J. Math. Anal. Appl.}\ }\textbf {\bibinfo {volume} {146}},\
%   \bibinfo {pages} {89} (\bibinfo {year} {1990})}\BibitemShut {NoStop}%
\bibitem [{\citenamefont {van Eijndhoven}\ and\ \citenamefont
  {Meyers}(1990)}]{SLLvanEijndhoven90}%
  \BibitemOpen
  \bibfield  {author} {\bibinfo {author} {\bibfnamefont {S.~J.~L.}\
  \bibnamefont {van Eijndhoven}}\ and\ \bibinfo {author} {\bibfnamefont
  {J.~L.~H.}\ \bibnamefont {Meyers}},\ }\href {\doibase
  https://doi.org/10.1016/0022-247X(90)90334-C} {\bibfield  {journal} {\bibinfo
   {journal} {J. Math. Anal. Appl.}\ }\textbf {\bibinfo {volume} {146}},\
  \bibinfo {pages} {89} (\bibinfo {year} {1990})}\BibitemShut {NoStop}%
\bibitem [{\citenamefont {G\'{o}rska}\ \emph {et~al.}(2019)\citenamefont
  {G\'{o}rska}, \citenamefont {Horzela},\ and\ \citenamefont
  {Szafraniec}}]{Gorska2018}%
  \BibitemOpen
  \bibfield  {author} {\bibinfo {author} {\bibfnamefont {K.}~\bibnamefont
  {G\'{o}rska}}, \bibinfo {author} {\bibfnamefont {A.}~\bibnamefont {Horzela}},
  \ and\ \bibinfo {author} {\bibfnamefont {F.~H.}\ \bibnamefont {Szafraniec}},\
  }\href {\doibase https://doi.org/10.1016/j.jmaa.2018.10.024} {\bibfield
  {journal} {\bibinfo  {journal} {J. Math. Appl. Anal.}\ }\textbf {\bibinfo
  {volume} {470}},\ \bibinfo {pages} {750} (\bibinfo {year}
  {2019})}\BibitemShut {NoStop}%
\bibitem [{\citenamefont
  {de~Gosson}(2021{\natexlab{a}})}]{Gosson_LMP111p73y2021}%
  \BibitemOpen
  \bibfield  {author} {\bibinfo {author} {\bibfnamefont {M.~A.}\ \bibnamefont
  {de~Gosson}},\ }\href {\doibase 10.1007/s11005-021-01410-4} {\bibfield
  {journal} {\bibinfo  {journal} {Lett. Math. Phys.}\ }\textbf {\bibinfo
  {volume} {111}},\ \bibinfo {pages} {73} (\bibinfo {year}
  {2021}{\natexlab{a}})}\BibitemShut {NoStop}%
\bibitem [{\citenamefont {Simon}(2000)}]{Simon_PRL84p2726y2000}%
  \BibitemOpen
  \bibfield  {author} {\bibinfo {author} {\bibfnamefont {R.}~\bibnamefont
  {Simon}},\ }\href {\doibase 10.1103/PhysRevLett.84.2726} {\bibfield
  {journal} {\bibinfo  {journal} {Phys. Rev. Lett.}\ }\textbf {\bibinfo
  {volume} {84}},\ \bibinfo {pages} {2726} (\bibinfo {year}
  {2000})}\BibitemShut {NoStop}%
\bibitem [{\citenamefont {de~Gosson}(2021{\natexlab{b}})}]{Gosson_QHA2021}%
  \BibitemOpen
  \bibfield  {author} {\bibinfo {author} {\bibfnamefont {M.~A.}\ \bibnamefont
  {de~Gosson}},\ }\href {\doibase doi:10.1515/9783110722772} {\emph {\bibinfo
  {title} {Quantum Harmonic Analysis. {A}n Introduction}}}\ (\bibinfo
  {publisher} {De Gruyter},\ \bibinfo {address} {Berlin, Boston},\ \bibinfo
  {year} {2021})\BibitemShut {NoStop}%
% \bibitem [{\citenamefont {G.}\ \emph {et~al.}(2004)\citenamefont {G.},
%   \citenamefont {A.},\ and\ \citenamefont {F.}}]{Adesso2004}%
%   \BibitemOpen
%   \bibfield  {author} {\bibinfo {author} {\bibfnamefont {A.}~\bibnamefont
%   {G.}}, \bibinfo {author} {\bibfnamefont {S.}~\bibnamefont {A.}}, \ and\
%   \bibinfo {author} {\bibfnamefont {I.}~\bibnamefont {F.}},\ }\href {\doibase
%   10.1103/PhysRevA.70.022318} {\bibfield  {journal} {\bibinfo  {journal} {Phys.
%   Rev. A}\ }\textbf {\bibinfo {volume} {70}},\ \bibinfo {pages} {022318}
%   (\bibinfo {year} {2004})}\BibitemShut {NoStop}% 
\bibitem [{\citenamefont {Adesso}\ \emph {et~al.}(2004)\citenamefont {Adesso},
  \citenamefont {Serafini},\ and\ \citenamefont {Illuminati}}]{Adesso2004}%
  \BibitemOpen
  \bibfield  {author} {\bibinfo {author} {\bibfnamefont {G.}~\bibnamefont
  {Adesso}}, \bibinfo {author} {\bibfnamefont {A.}~\bibnamefont {Serafini}}, \
  and\ \bibinfo {author} {\bibfnamefont {F.}~\bibnamefont {Illuminati}},\
  }\href {\doibase 10.1103/PhysRevA.70.022318} {\bibfield  {journal} {\bibinfo
  {journal} {Phys. Rev. A}\ }\textbf {\bibinfo {volume} {70}},\ \bibinfo
  {pages} {022318} (\bibinfo {year} {2004})}\BibitemShut {NoStop}%
\bibitem [{\citenamefont {Li}\ \emph {et~al.}(2023)\citenamefont {Li},
  \citenamefont {Das}, \citenamefont {Tserkis}, \citenamefont {Narang},
  \citenamefont {Lam},\ and\ \citenamefont {Assad}}]{Li_SR13p11722y2023}%
  \BibitemOpen
  \bibfield  {author} {\bibinfo {author} {\bibfnamefont {B.}~\bibnamefont
  {Li}}, \bibinfo {author} {\bibfnamefont {A.}~\bibnamefont {Das}}, \bibinfo
  {author} {\bibfnamefont {S.}~\bibnamefont {Tserkis}}, \bibinfo {author}
  {\bibfnamefont {P.}~\bibnamefont {Narang}}, \bibinfo {author} {\bibfnamefont
  {P.~K.}\ \bibnamefont {Lam}}, \ and\ \bibinfo {author} {\bibfnamefont
  {S.~M.}\ \bibnamefont {Assad}},\ }\href {\doibase 10.1038/s41598-023-38572-1}
  {\bibfield  {journal} {\bibinfo  {journal} {Sci. Rep.}\ }\textbf {\bibinfo
  {volume} {13}},\ \bibinfo {pages} {11722} (\bibinfo {year}
  {2023})}\BibitemShut {NoStop}%
\bibitem [{\citenamefont {Alonso-López}\ \emph {et~al.}(2023)\citenamefont
  {Alonso-López}, \citenamefont {Cembranos}, \citenamefont {Díaz-Guerra},\
  and\ \citenamefont {Mínguez-Sánchez}}]{Alonso_EPJD77p43y2023}%
  \BibitemOpen
  \bibfield  {author} {\bibinfo {author} {\bibfnamefont {D.}~\bibnamefont
  {Alonso-López}}, \bibinfo {author} {\bibfnamefont {J.~A.~R.}\ \bibnamefont
  {Cembranos}}, \bibinfo {author} {\bibfnamefont {D.}~\bibnamefont
  {Díaz-Guerra}}, \ and\ \bibinfo {author} {\bibfnamefont {A.}~\bibnamefont
  {Mínguez-Sánchez}},\ }\href {\doibase 10.1140/epjd/s10053-023-00629-1}
  {\bibfield  {journal} {\bibinfo  {journal} {Eur. Phys. J. D}\ }\textbf
  {\bibinfo {volume} {77}},\ \bibinfo {pages} {43} (\bibinfo {year}
  {2023})}\BibitemShut {NoStop}%
\bibitem [{\citenamefont {Szafraniec}(2004)}]{FHSbook}%
  \BibitemOpen
  \bibfield  {author} {\bibinfo {author} {\bibfnamefont {F.~H.}\ \bibnamefont
  {Szafraniec}},\ }\href
  {https://ruj.uj.edu.pl/xmlui/bitstream/handle/item/272364/szafraniec_przestrzenie_hilberta_z_jadrem_reprodukujacym_2004.pdf?sequence=1&isAllowed=y}
  {\emph {\bibinfo {title} {Przestrzenie {Hilberta} z j\c{a}drem
  reprodukuj\c{a}cym ({Reproducing} kernel {Hilbert} space)}}}\ (\bibinfo
  {publisher} {Wydawnictwo Uniwersytetu Jagiello\'{n}skiego},\ \bibinfo {year}
  {2004})\BibitemShut {NoStop}%
% \bibitem [{\citenamefont {Twareque~Ali}\ \emph {et~al.}(2014)\citenamefont
%   {Twareque~Ali}, \citenamefont {G\'{o}rska}, \citenamefont {Horzela},\ and\
%   \citenamefont {Szafraniec}}]{STAli2014}%
%   \BibitemOpen
%   \bibfield  {author} {\bibinfo {author} {\bibfnamefont {S.}~\bibnamefont
%   {Twareque~Ali}}, \bibinfo {author} {\bibfnamefont {K.}~\bibnamefont
%   {G\'{o}rska}}, \bibinfo {author} {\bibfnamefont {A.}~\bibnamefont {Horzela}},
%   \ and\ \bibinfo {author} {\bibfnamefont {F.~H.}\ \bibnamefont {Szafraniec}},\
%   }\href {\doibase https://doi.org/10.1063/1.4861932} {\bibfield  {journal}
%   {\bibinfo  {journal} {J. Math. Phys.}\ }\textbf {\bibinfo {volume} {55}},\
%   \bibinfo {pages} {012107} (\bibinfo {year} {2014})}\BibitemShut {NoStop}%
\bibitem [{\citenamefont {Ali}\ \emph {et~al.}(2014)\citenamefont {Ali},
  \citenamefont {G\'{o}rska}, \citenamefont {Horzela},\ and\ \citenamefont
  {Szafraniec}}]{STAli2014}%
  \BibitemOpen
  \bibfield  {author} {\bibinfo {author} {\bibfnamefont {T.~S.}\ \bibnamefont
  {Ali}}, \bibinfo {author} {\bibfnamefont {K.}~\bibnamefont {G\'{o}rska}},
  \bibinfo {author} {\bibfnamefont {A.}~\bibnamefont {Horzela}}, \ and\
  \bibinfo {author} {\bibfnamefont {F.~H.}\ \bibnamefont {Szafraniec}},\ }\href
  {\doibase https://doi.org/10.1063/1.4861932} {\bibfield  {journal} {\bibinfo
  {journal} {J. Math. Phys.}\ }\textbf {\bibinfo {volume} {55}},\ \bibinfo
  {pages} {012107} (\bibinfo {year} {2014})}\BibitemShut {NoStop}%
\bibitem [{\citenamefont {Ismail}(2016)}]{MEHIsmail16}%
  \BibitemOpen
  \bibfield  {author} {\bibinfo {author} {\bibfnamefont {M.~E.~H.}\
  \bibnamefont {Ismail}},\ }\href {\doibase https://doi.org/10.1090/tran/6358}
  {\bibfield  {journal} {\bibinfo  {journal} {Trans. Am. Math. Soc.}\ }\textbf
  {\bibinfo {volume} {368}},\ \bibinfo {pages} {1189} (\bibinfo {year}
  {2016})}\BibitemShut {NoStop}%
\bibitem [{\citenamefont {W\"{u}nsche}(2015)}]{AWunsche15}%
  \BibitemOpen
  \bibfield  {author} {\bibinfo {author} {\bibfnamefont {A.}~\bibnamefont
  {W\"{u}nsche}},\ }\href {\doibase 10.4236/am.2015.612188} {\bibfield
  {journal} {\bibinfo  {journal} {Appl. Math.}\ }\textbf {\bibinfo {volume}
  {6}},\ \bibinfo {pages} {2142} (\bibinfo {year} {2015})}\BibitemShut
  {NoStop}%
\bibitem [{\citenamefont {Ghanmi}(2008)}]{ghanmi1}%
  \BibitemOpen
  \bibfield  {author} {\bibinfo {author} {\bibfnamefont {A.}~\bibnamefont
  {Ghanmi}},\ }\href {\doibase https://doi.org/10.1016/j.jmaa.2007.10.001}
  {\bibfield  {journal} {\bibinfo  {journal} {J. Math. Anal. Appl.}\ }\textbf
  {\bibinfo {volume} {340}},\ \bibinfo {pages} {1395} (\bibinfo {year}
  {2008})}\BibitemShut {NoStop}%
\bibitem [{\citenamefont {Ghanmi}(2013)}]{ghanmi2}%
  \BibitemOpen
  \bibfield  {author} {\bibinfo {author} {\bibfnamefont {A.}~\bibnamefont
  {Ghanmi}},\ }\href {\doibase https://doi.org/10.1080/10652469.2013.772172}
  {\bibfield  {journal} {\bibinfo  {journal} {Integr. Transf. Spec. F.}\
  }\textbf {\bibinfo {volume} {24}},\ \bibinfo {pages} {884} (\bibinfo {year}
  {2013})}\BibitemShut {NoStop}%
\bibitem [{\citenamefont {It\^{o}}(1952)}]{KIto52}%
  \BibitemOpen
  \bibfield  {author} {\bibinfo {author} {\bibfnamefont {K.}~\bibnamefont
  {It\^{o}}},\ }\href {\doibase 10.4099/jjm1924.22.0_63} {\bibfield  {journal}
  {\bibinfo  {journal} {Jpn. J. Math.}\ }\textbf {\bibinfo {volume} {22}},\
  \bibinfo {pages} {63} (\bibinfo {year} {1952})}\BibitemShut {NoStop}%
\bibitem [{\citenamefont {Cotfas}\ \emph {et~al.}(2010)\citenamefont {Cotfas},
  \citenamefont {Gazeau},\ and\ \citenamefont {G\'{o}rska}}]{NCotfas10}%
  \BibitemOpen
  \bibfield  {author} {\bibinfo {author} {\bibfnamefont {N.}~\bibnamefont
  {Cotfas}}, \bibinfo {author} {\bibfnamefont {J.-P.}\ \bibnamefont {Gazeau}},
  \ and\ \bibinfo {author} {\bibfnamefont {K.}~\bibnamefont {G\'{o}rska}},\
  }\href {\doibase 10.1088/1751-8113/43/30/305304} {\bibfield  {journal}
  {\bibinfo  {journal} {J. Phys. A: Math. Theor.}\ }\textbf {\bibinfo {volume}
  {43}},\ \bibinfo {pages} {305304 (14pp)} (\bibinfo {year}
  {2010})}\BibitemShut {NoStop}%
\bibitem [{\citenamefont {Hong-{Yi}}\ and\ \citenamefont
  {Klauder}(1994)}]{Fan_PRA49p704y1994}%
  \BibitemOpen
  \bibfield  {author} {\bibinfo {author} {\bibfnamefont {F.}~\bibnamefont
  {Hong-{Yi}}}\ and\ \bibinfo {author} {\bibfnamefont {J.~R.}\ \bibnamefont
  {Klauder}},\ }\href {\doibase 10.1103/PhysRevA.49.704} {\bibfield  {journal}
  {\bibinfo  {journal} {Phys. Rev. A}\ }\textbf {\bibinfo {volume} {49}},\
  \bibinfo {pages} {704} (\bibinfo {year} {1994})}\BibitemShut {NoStop}%
\bibitem [{\citenamefont {Wigner}(1932)}]{Wigner_PR40p749y1932}%
  \BibitemOpen
  \bibfield  {author} {\bibinfo {author} {\bibfnamefont {E.}~\bibnamefont
  {Wigner}},\ }\href {\doibase 10.1103/PhysRev.40.749} {\bibfield  {journal}
  {\bibinfo  {journal} {Phys. Rev.}\ }\textbf {\bibinfo {volume} {40}},\
  \bibinfo {pages} {749} (\bibinfo {year} {1932})}\BibitemShut {NoStop}%
\bibitem [{\citenamefont {de~Gosson}(2017)}]{Gosson2017}%
  \BibitemOpen
  \bibfield  {author} {\bibinfo {author} {\bibfnamefont {M.}~\bibnamefont
  {de~Gosson}},\ }\href {\doibase 10.1142/q0089} {\emph {\bibinfo {title} {The
  {Wigner} Transform}}}\ (\bibinfo  {publisher} {World Scientific (Europe)},\
  \bibinfo {year} {2017})\BibitemShut {NoStop}%
\bibitem [{\citenamefont {Peres}(1996)}]{Peres_PRL77p1413y1996}%
  \BibitemOpen
  \bibfield  {author} {\bibinfo {author} {\bibfnamefont {A.}~\bibnamefont
  {Peres}},\ }\href {\doibase 10.1103/PhysRevLett.77.1413} {\bibfield
  {journal} {\bibinfo  {journal} {Phys. Rev. Lett.}\ }\textbf {\bibinfo
  {volume} {77}},\ \bibinfo {pages} {1413} (\bibinfo {year}
  {1996})}\BibitemShut {NoStop}%
\bibitem [{\citenamefont {Horodecki}\ \emph {et~al.}(1996)\citenamefont
  {Horodecki}, \citenamefont {Horodecki},\ and\ \citenamefont
  {Horodecki}}]{Horodecki_PLA223p1y1996}%
  \BibitemOpen
  \bibfield  {author} {\bibinfo {author} {\bibfnamefont {M.}~\bibnamefont
  {Horodecki}}, \bibinfo {author} {\bibfnamefont {P.}~\bibnamefont
  {Horodecki}}, \ and\ \bibinfo {author} {\bibfnamefont {R.}~\bibnamefont
  {Horodecki}},\ }\href {\doibase
  https://doi.org/10.1016/S0375-9601(96)00706-2} {\bibfield  {journal}
  {\bibinfo  {journal} {Phys. Lett. A}\ }\textbf {\bibinfo {volume} {223}},\
  \bibinfo {pages} {1} (\bibinfo {year} {1996})}\BibitemShut {NoStop}%
\bibitem [{\citenamefont {Vidal}\ and\ \citenamefont
  {Werner}(2002)}]{Vidal_PRA65p032314y2002}%
  \BibitemOpen
  \bibfield  {author} {\bibinfo {author} {\bibfnamefont {G.}~\bibnamefont
  {Vidal}}\ and\ \bibinfo {author} {\bibfnamefont {R.~F.}\ \bibnamefont
  {Werner}},\ }\href {\doibase 10.1103/PhysRevA.65.032314} {\bibfield
  {journal} {\bibinfo  {journal} {Phys. Rev. A}\ }\textbf {\bibinfo {volume}
  {65}},\ \bibinfo {pages} {032314} (\bibinfo {year} {2002})}\BibitemShut
  {NoStop}%
\bibitem [{\citenamefont {Plenio}(2005)}]{Plenio_PRL95p090503y2005}%
  \BibitemOpen
  \bibfield  {author} {\bibinfo {author} {\bibfnamefont {M.~B.}\ \bibnamefont
  {Plenio}},\ }\href {\doibase 10.1103/PhysRevLett.95.090503} {\bibfield
  {journal} {\bibinfo  {journal} {Phys. Rev. Lett.}\ }\textbf {\bibinfo
  {volume} {95}},\ \bibinfo {pages} {090503} (\bibinfo {year}
  {2005})}\BibitemShut {NoStop}%
\bibitem [{\citenamefont {Kao}\ and\ \citenamefont
  {Chou}(2018)}]{Kao_SR8p7394y2018}%
  \BibitemOpen
  \bibfield  {author} {\bibinfo {author} {\bibfnamefont {J.}~\bibnamefont
  {Kao}}\ and\ \bibinfo {author} {\bibfnamefont {C.}~\bibnamefont {Chou}},\
  }\href {\doibase 10.1038/s41598-018-25781-2} {\bibfield  {journal} {\bibinfo
  {journal} {Sci. Rep.}\ }\textbf {\bibinfo {volume} {8}},\ \bibinfo {pages}
  {7394} (\bibinfo {year} {2018})}\BibitemShut {NoStop}%
\bibitem [{\citenamefont {Gonzalez-Henao}\ \emph {et~al.}(2015)\citenamefont
  {Gonzalez-Henao}, \citenamefont {Pugliese}, \citenamefont {Euzzor},
  \citenamefont {Abdalah}, \citenamefont {Meucci},\ and\ \citenamefont
  {Roversi}}]{Gonzalez_SR5p1315y2015}%
  \BibitemOpen
  \bibfield  {author} {\bibinfo {author} {\bibfnamefont {J.~C.}\ \bibnamefont
  {Gonzalez-Henao}}, \bibinfo {author} {\bibfnamefont {E.}~\bibnamefont
  {Pugliese}}, \bibinfo {author} {\bibfnamefont {S.}~\bibnamefont {Euzzor}},
  \bibinfo {author} {\bibfnamefont {S.}~\bibnamefont {Abdalah}}, \bibinfo
  {author} {\bibfnamefont {R.}~\bibnamefont {Meucci}}, \ and\ \bibinfo {author}
  {\bibfnamefont {J.~A.}\ \bibnamefont {Roversi}},\ }\href {\doibase
  10.1038/srep13152} {\bibfield  {journal} {\bibinfo  {journal} {Sci. Rep.}\
  }\textbf {\bibinfo {volume} {5}},\ \bibinfo {pages} {13152} (\bibinfo {year}
  {2015})}\BibitemShut {NoStop}%
\bibitem [{\citenamefont {Makarov}(2018)}]{Makarov_PRE97p042203y2018}%
  \BibitemOpen
  \bibfield  {author} {\bibinfo {author} {\bibfnamefont {D.~N.}\ \bibnamefont
  {Makarov}},\ }\href {\doibase 10.1103/PhysRevE.97.042203} {\bibfield
  {journal} {\bibinfo  {journal} {Phys. Rev. E}\ }\textbf {\bibinfo {volume}
  {97}},\ \bibinfo {pages} {042203} (\bibinfo {year} {2018})}\BibitemShut
  {NoStop}%
\bibitem [{\citenamefont {Roman}\ and\ \citenamefont {Rota}(1978)}]{RomanS}%
  \BibitemOpen
  \bibfield  {author} {\bibinfo {author} {\bibfnamefont {S.~M.}\ \bibnamefont
  {Roman}}\ and\ \bibinfo {author} {\bibfnamefont {G.-C.}\ \bibnamefont
  {Rota}},\ }\href {\doibase https://doi.org/10.1016/0001-8708(78)90087-7}
  {\bibfield  {journal} {\bibinfo  {journal} {Adv. Math.}\ }\textbf {\bibinfo
  {volume} {27}},\ \bibinfo {pages} {95} (\bibinfo {year} {1978})}\BibitemShut
  {NoStop}%
\end{thebibliography}
%

\section*{Acknowledgements}
The research project is partially supported by a subsidy
from the Polish Ministry of Science and Higher Education
and the program 'Initiative for Excellence – Research University' for the AGH University of Krakow. 
\end{document}